\documentclass[aps,prc,a4paper,nofootinbib,showpacs,showkeys,
preprintnumbers,superscriptaddress,twocolumn]{revtex4}

\usepackage{amsmath}
\usepackage{amssymb}
\usepackage{bm}
\usepackage{graphicx}
\usepackage{color}

\begin{document}

\title{Modelling Early Stages of Relativistic Heavy Ion Collisions:\\
Coupling Relativistic Transport Theory to Decaying Color-electric Flux Tubes }

\author{M. Ruggieri}
\affiliation{Department of Physics and Astronomy, University of Catania, Via S. Sofia 64, I-95125 Catania}

\author{A. Puglisi}
\affiliation{Department of Physics and Astronomy, University of Catania, Via S. Sofia 64, I-95125 Catania}
\affiliation{INFN-Laboratori Nazionali del Sud, Via S. Sofia 62, I-95123 Catania, Italy}

\author{L. Oliva}
\affiliation{Department of Physics and Astronomy, University of Catania, Via S. Sofia 64, I-95125 Catania}
\affiliation{INFN-Laboratori Nazionali del Sud, Via S. Sofia 62, I-95123 Catania, Italy}

\author{S. Plumari}
\affiliation{Department of Physics and Astronomy, University of Catania, Via S. Sofia 64, I-95125 Catania}
\affiliation{INFN-Laboratori Nazionali del Sud, Via S. Sofia 62, I-95123 Catania, Italy}

\author{F. Scardina}
\affiliation{Department of Physics and Astronomy, University of Catania, Via S. Sofia 64, I-95125 Catania}
\affiliation{INFN-Laboratori Nazionali del Sud, Via S. Sofia 62, I-95123 Catania, Italy}

\author{V. Greco}
\affiliation{Department of Physics and Astronomy, University of Catania, Via S. Sofia 64, I-95125 Catania}
\affiliation{INFN-Laboratori Nazionali del Sud, Via S. Sofia 62, I-95123 Catania, Italy}


\begin{abstract}
In this study we 
model early times dynamics of relativistic heavy ion collisions by an initial color electric field
which then decays to a plasma by the Schwinger mechanism,
coupling the dynamical evolution of the initial color field
to the dynamics of the many particles system produced by the decay. 
The latter
is described by relativistic kinetic theory in which we fix the ratio 
$\eta/s$ rather than insisting on specific microscopic processes,
and the backreaction on the color field is taken into account by solving
self-consistently the kinetic and the field equations.
We study isotropization and thermalization of the system produced by the field decay
for a static box and for a $1+1$D expanding geometry.
We find that regardless of the viscosity of the produced plasma,
the initial color electric field decays within $1$ fm/c;
however in the case $\eta/s$ is large, oscillations of the field
are effective along all the entire time evolution of the system,
which affect the late times evolution of the ratio between longitudinal
and transverse pressure. In case of small $\eta/s$  ($\eta/s\lesssim0.3$)
we find $\tau_{isotropization}\approx 0.8$ fm/c
and $\tau_{thermalization}\approx 1$ fm/c in agreement with the common lore of hydrodynamics.
Moreover we have investigated the effect of 
turning from the relaxation time approximation to the Chapman-Enskog one:
we find that this improvement affects mainly the early times evolution
of the physical quantities, the effect being milder in the late times
evolution.
\end{abstract}


\maketitle

\section{Introduction}
The understanding of early times dynamics is one of the most interesting and compelling 
problems of heavy ion collisions at
ultrarelativistic energy. According to the standard picture
of such collision processes, immediately after the collision
an ensemble of strong longitudinal color electric and color magnetic
coherent fields known as Glasma is produced
\cite{Lappi:2006fp,Gelis:2006dv,Lappi:2008eq,Gelis:2009wh,Fukushima:2011nq,Iida:2013qwa}.
Inhomogeneities in the Glasma fields are amplified by 
plasma instabilities, which eventually lead to loss of coherence of the color fields
and to a locally isotropized particle plasma
\cite{Romatschke:2006nk,Fujii:2009kb,Gelis:2013rba,Randrup:2003cw,Mrowczynski:2005ki,Mrowczynski:2006ad,
Attems:2012js,Rebhan:2008uj,
Schenke:2006xu,Arnold:2003rq,Arnold:2004ti,Arnold:2005vb,Arnold:2005ef,Rebhan:2004ur,Rebhan:2005re}. 
Recently the early isotropization problem has been studied also by means of 
AdS/CFT methods  \cite{Bellantuono:2015hxa,vanderSchee:2013pia,Kovchegov:2007pq,Heller:2011ju,
Heller:2014wfa}.

Besides plasma instabilities, a mechanism responsible for the initial field
decays is the one introduced by Schwinger in the context of Quantum Electrodynamics 
\cite{Heisenberg:1935qt,Schwinger:1951nm}, 
known as the Schwinger effect
which consists in a vacuum instability
towards the creation of particle pairs by a strong electric field,
and it is related to the existence of an imaginary part in the quantum 
effective action of a pure electric field, see \cite{Dunne:2004nc} for a review.
The problem of pair formation in strong electric
fields has been considered recently by means of real time lattice 
simulations \cite{Hebenstreit:2013baa,Hebenstreit:2013qxa,Gelis:2013oca}.
Moreover non abelian generalizations of the Schwinger production rate has been found
for the case of both static and time dependent 
field~\cite{Nayak:2007ds,Nayak:2007ju,Cooper:2006ag,Cooper:2006mt,Cooper:2005rk,Nayak:2005pf}.
In the context of quark-gluon plasma physics it has been considered as a mechanism for
the color field decays in \cite{Casher:1978wy,Glendenning:1983qq,Bialas:1984wv,Bialas:1984ap,Bialas:1985is,
Gyulassy:1986jq,Gatoff:1987uf,Elze:1986qd,Elze:1986hq,Bialas:1986mt,
Florkowski:2003mm,Bajan:2001fs,Bialas:1989hc,Dyrek:1988eb,
Bialas:1987en,Ryblewski:2013eja,Florkowski:2014ska,Florkowski:2010zz,
Tanji:2011di,Tanji:2008ku,Tanji:2010ui}.

The purpose of the present study is to model early times dynamics of the system
produced in relativistic heavy ion collisions by an initial color electric field
which then decays to a plasma by the Schwinger effect.
The focus of our study is to compute quantities which
serve as indicators of thermalization and isotropization of the plasma.
In particular, we couple the dynamical evolution of the color field
to the dynamics of the many particles system produced by the decay, the latter
being described by relativistic kinetic theory
\cite{Plumari:2012xz,Plumari:2012ep,Ruggieri:2013bda,Ruggieri:2013ova}.
As in the previous studies
on this subject we assume that the dynamics of the color field is abelian,
hence it satisfies the classical field Maxwell equations,
see \cite{Florkowski:2010zz} for a review.
Moreover in the present approach we neglect for simplicity 
initial longitudinal color magnetic fields which are present in the more
complete description of the Glasma state.
We consider two simple cases: the first one is a system
evolving in a static box; then we consider a system with a longitudinal
expansion which has a greater physical interest than the static box
because it is closer to the picture of the early times dynamics of
relativistic heavy ion collisions.

Besides being the first study in which a Monte Carlo method
is used to simulate the Schwinger effect in the context of early times dynamics of high energy collisions,
we improve previous studies which mainly rely on 
Relaxation Time Approximation (RTA) \cite{Ryblewski:2013eja,Florkowski:2014ska}
or  on a linearization of the conductive electric current \cite{Gatoff:1987uf},
by avoiding any ansatz both on the electric current and on the collision integral in the Boltzmann equation
which permits to go beyond the RTA.
The approach developed here, based on a stochastic solution of the relativistic
Boltzman equation, has in perspective the advantage to be easily extended to 
$3+1$D realistic simulations, allowing to study the possible impact of the early dynamics on observables
like radial and elliptic flow~\cite{prepa}.

In both the cases of the static box and of the longitudinal expanding system
we have focused on the calculation of $P_L/P_T$, where $P_L$ and $P_T$ 
denote the longitudinal and transverse pressures respectively.
For the particular initial field configuration used in this work,
namely an electric field along the longitudinal direction,
one has $P_L/P_T = -1$; however the field decays onto particles thanks to the Schwinger effect and
particle dynamics removes this initial anisotropy,
the efficiency of isotropization being related to the coupling among
the particles. In our simulations we fix the ratio $\eta/s$ where $\eta$
is the shear viscosity and $s$ the entropy density, and compute
locally the corresponding cross section by means of the Chapman-Enskog (CE) 
approximation \cite{Plumari:2012xz,Plumari:2012ep}. We find that
increasing $\eta/s$ the system is less efficient in removing the initial
anisotropy in agreement with previous studies, 
although regardless of $\eta/s$ the longitudinal pressure becomes
positive within a small fraction of fm/c, $\tau\approx 0.1$ fm/c; moreover for large values of $\eta/s$ 
we find persistent oscillations of the color field which affects the evolution
of physical quantities like $P_L/P_T$ and the proper energy density,
in agreement with \cite{Ryblewski:2013eja,Florkowski:2014ska}.
Even if we go beyond the RTA we find that only the early times evolution
is affected, leaving the main behaviours 
very mildly affected.

The plan of the article is as follows. In Section II we review briefly the 
abelian flux tube model which we implement in our simulations.
In Section III we review relativistic transport theory.
In Section IV we present our results of the simulations for the case 
of a static box. In Section V we discuss in detail our results
for the case of a box with a longitudinal expansion. Finally in Section VI we
draw our conclusions.

\section{Abelian Flux Tube Model}
In this Section we briefly summarize the abelian flux tube model (AFTm)
\cite{Casher:1978wy,Glendenning:1983qq,Bialas:1984wv,Bialas:1984ap,Bialas:1985is,
Gyulassy:1986jq,Gatoff:1987uf,Elze:1986qd,Elze:1986hq,Bialas:1986mt,
Florkowski:2003mm,Bajan:2001fs,Bialas:1989hc,Dyrek:1988eb,
Bialas:1987en,Ryblewski:2013eja,Florkowski:2014ska,Florkowski:2010zz} 
which we implement in our simulations
of the initial stage of relativistic heavy ion collisions.
In the present study we do not insist on implementing the most realistic geometrical condition
relevant for heavy ion collisions experiments, in which one should take into account 
several flux tubes in the transverse plane similar to the glasma configuration; 
rather we consider a simpler situation 
in which there is only one flux tube of a given transverse area, and study its 
dynamical evolution by coupling the field equations to relativistic kinetic theory
for the particle quanta produced by the decay of the field itself,
leaving to upcoming works the study of more realistic  
initial conditions.

The main assumptions of the AFTm 
are:
\begin{itemize}
\item In the initial condition a color electric field is present,
which is produced by color charges in the colliding nuclei; 
\item The color electric field decays into particle quanta by the Schwinger mechanism;
\item The particle quanta  
propagate in the medium colliding and interacting with the background of the color field;
\item Particle creation as well as particle currents affect in a self-consistent way
the color electric field;
\item The field dynamics is abelian, namely it satisfies the abelian Maxwell equations.
\end{itemize}
The last assumption is quite strong because initial gauge fields are quite large, 
nevertheless it permits the easiest implementation of the coupling of
classical field equations to relativistic kinetic theory in presence of the Schwinger mechanism,
hence we will limit ourselves to consider abelian dynamics of the classical fields
leaving the introduction of non abelian field dynamics to future works.

We will discuss more about kinetic theory in the next Section, therefore in this Section
we mainly focus on the  Schwinger mechanism which is responsible for particle
production in the early stage of the collision, and on the Monte Carlo implementation we use
to simulate the Schwinger effect.

In this Section we closely follow \cite{Ryblewski:2013eja} at the same time
adopting a slightly different notation which is more convenient for this article. 
Moreover, since the very beginning we assume the initial color field
to be polarized along the $3^{rd}$ direction of adjoint color space,
meaning that only one particular color charge is present in the 
two colliding nuclei~\cite{Ryblewski:2013eja}.
Moreover we assume only one component of the electric field is present,
namely the one in the longitudinal direction, which we denote by $E$ in the following.
The latter is a consequence of the fact that in the initial condition
we assume the field is purely longitudinal, and transverse currents are not produced during time evolution
of the system if transverse expansion does not take place implying a vanishing
transverse field.
The number of  pairs per unit of spacetime
and invariant momentum space produced by the decay of the electric field by the Schwinger effect is
(we assume massles quanta throughout this article)
\begin{equation}
\frac{dN_{jc}}{d\Gamma}\equiv
p_0\frac{dN_{jc}}{d^4x d^2 p_T dp_z} = {\cal R}_{jc}(p_T)
\delta(p_z)p_0~,
\label{eq:SF}
\end{equation}
with
\begin{equation}
{\cal R}_{jc}(p_T) =\frac{{\cal E}_{jc}}{4\pi^3}
\left|
\ln
\left(
1\pm e^{-\pi p_T^2/{\cal E}_{jc}}
\right)
\right|~,
\label{eq:RpT}
\end{equation}
the plus (minus) sign corresponding to the creation of a boson (fermion-antifermion) pair.
In this equation $p_T$, $p_z$  refer to
each of the two particles created by the tunneling process;
${\cal E}_{jc}$ is the effective force which acts on the tunneling pair and it depends
on color and flavor; it can be written as
\begin{equation}
{\cal E}_{jc} = \left(
g|Q_{jc}E| - \sigma_{j}
\right)
\theta
\left(
g|Q_{jc}E| - \sigma_{j}
\right)~,
\label{eq:sigma}
\end{equation}
where $\sigma_j$ denotes the string tension depending on the kind of flavor considered.
Moreover $p_0=\sqrt{p_T^2 + p_z^2}$ corresponds to the single particle kinetic energy.

The $Q_{jc}$ are color-flavor charges which, in the case of quarks,
correspond to the eigenvalues of the $T_3$ operator:
\begin{equation}
Q_{j1}=\frac{1}{2}~,~~~~Q_{j2}=-\frac{1}{2}~,~~~~Q_{j3}=0~,~~~~~j=1,N_f;
\end{equation} 
for antiquarks, corresponding to negative values of $j$, the color-flavor charges
are just minus the corresponding charges for quarks. 
Finally for gluons (which 
in our notation correspond to $j=0$) the charges are obtained
by building gluons up as the octet of the $3\otimes \bar 3$
in color space; in particular
\begin{equation}
Q_{01}=1~,~~~~Q_{02}=\frac{1}{2}~,~~~~Q_{03}=-\frac{1}{2}~,
\end{equation} 
and $Q_{04}=-Q_{01}$, $Q_{05}=-Q_{02}$, $Q_{06}=-Q_{03}$.
The above charges are obtained easily \cite{Florkowski:2010zz}: for example, the gluon we label with subscript $1$
can be represented, from the point of view of transformations in color space, as a red-antigreen state; 
its $T_3$ charge is then given by difference of $T_3$ charges of a red and a green state.
We notice that we have only six gluons out of eight, corresponding to the non-diagonal color generators;
the gluon fields corresponding to the diagonal color generators have vanishing coupling with the background
field, hence they cannot be produced by the Schwinger effect.

As a final remark we would like to observe that, even if the model is named abelian, such nomenclature
simply refers to the fact that in the evolution equation for the classical field, self-interacion terms
coming from non vanishing structure constants of the color gauge group are neglected \cite{Florkowski:2010zz}. 
However interactions among the classical field and gluons are still present in this calculations,
thanks both to the Schwinger effect which produces charged gluons, and to conduction currents which
affect the evolution of the field, see the next Section for more details.

\section{Relativistic Transport Theory Coupled to Maxwell Equations}

Our calculation scheme is based on the Relativistic Transport Boltzmann equation which, in the presence of 
a gauge field $F^{\mu\nu}$, can be written as follows:
\begin{equation}
\label{Boltzmann}
\left( p^{\mu}\partial_{\mu} + g Q_{jc} F^{\mu\nu}p_{\nu}\partial_{\mu}^{p} \right)f_{jc}(x,p)=\frac{dN_{jc}}{d\Gamma}+{\cal C}_{jc}[f]
\end{equation}
where $f_{jc}(x,p)$ is the distribution function for flavour $j$ and color $c$, $F^{\mu\nu}$ is the electromagnetic tensor. On the right hand side we have the source term $dN/d\Gamma$ which describes the creation of quarks, antiquarks and gluons due to the decay of the color electric field and ${\cal C}[f]$ which represents the collision integral.
Considering only $2\to 2$ body elastic scatterings, the collision integral can be written as:
\begin{eqnarray}
{\cal C}[f]&=&
 \int\frac{d^3p_2}{2E_2(2\pi)^3}  \frac{d^3p_{1^\prime}}{2E_{1^\prime}(2\pi)^3}  \frac{d^3p_2^\prime}{2E_2^\prime(2\pi)^3}\,
(f_{1^\prime}f_{2^\prime} - f_1 f_2)\nonumber\\ 
&&\times|{\cal M}|^2\delta^4(p_1 + p_2 - p_{1^\prime} - p_{2^\prime})~,
\end{eqnarray}
where we omit flavour and color indices for simplicity,  ${\cal M}$ is the transition matrix for the elastic process linked to the differential cross section through $|{\cal M}|^2=16\pi s^2d\sigma/dt$, being $s$ the Mandelstam variable.

In our simulations we solve numerically Eq. (\ref{Boltzmann}) 
using the test particles method and the collision integral is computed 
using Monte Carlo methods based on the stochastic interpretation 
of transition amplitude \cite{Xu:2008av,Xu:2007jv,Bratkovskaya:2011wp,Ferini:2008he,
Plumari_BARI,Plumari:2012xz,Ruggieri:2013bda,Ruggieri:2013ova}.

The evolution of the electric field is given by the Maxwell equations:
\begin{equation}
\frac{\partial E}{\partial z} = \rho~,~~~
\frac{\partial E}{\partial t} = -j~,
\label{eq:MAX_tt}
\end{equation}
where $\rho$ corresponds to the color charge density,
\begin{equation}
\rho = g\sum_{j,c}Q_{jc}\int d^3\bm p  f_{jc}(p)~,
\end{equation}
where $j,c$ denote flavor and color respectively; the sum in the above equation runs over quarks, antiquarks and gluons.
On the other hand $j$ corresponds to the color electric current
which is given by the sum of two contributions:
in fact the Schwinger effect can be described as a dielectric breakdown in which
dipoles are produced by quantum tunneling hence changing the local dipole moment of the vacuum,
and the charges move in the medium due to the residual electric field giving rise to a conductive current.
Following~\cite{Ryblewski:2013eja} we name displacement current, $j_D$, and
matter current, $j_M$:
\begin{equation}
j = j_M + j_D~.
\end{equation}
Here $j_M$ is a colored generalization of the usual electric current density
which in a continuum notation is given by
\begin{equation}
j_M^\mu = g\sum_{j,c}Q_{jc}\int\frac{d^3\bm p}{p_0}p^\mu f_{jc}(p)~.
\end{equation}
The displacement current arises from the polarization of the vacuum due to the decay
of the electric field by the Schwinger mechanism: more precisely it is given by the
time derivative of the local dipole moment induced by the particles pop-up,
in the same way a time variation of the local dipole moment in a medium gives rise
to a change in the local electric field \cite{Landau}.
According to the quantum tunneling interpretation of the Schwinger effect~\cite{Casher:1978wy}
the dipole moment can be computed as $2p_0/E$ where $p_0$ corresponds to 
the kinetic energy
of the particles coming out from the vacuum; taking into account Eq.~\eqref{eq:SF}
$j_D$ can be written, in the reference frame where particles are produced with vanishing
longitudinal momentum, as
\begin{equation}
j_D =\sum_{j=0}^{N_f}\sum_{c=1}^3
\int\frac{d^3\bm p}{p_0}
 \frac{dN_{jc}}{d\Gamma} 
\frac{2p_T}{E}~,
\label{eq:jD}
\end{equation}
where $N_f$ corresponds to the number of flavors in the calculation.
The color charge and current densities depend on the particle distribution
function: hence they link the Maxwell equations~\eqref{eq:MAX_tt} to the kinetic 
equation~\eqref{Boltzmann}. 
We solve self-consistently the field and kinetic equations:
in this way we take into account the back reaction of particle production and propagation
on the color field.

At variance with the standard use of transport theory, in which one fixes a set of microscopic processes into the collision integral, we have developed an approach that
fixes the total cross section in order to have the wanted $\eta/s$. By means of this scheme we are able to use the Boltzmann equation to simulate the dynamical evolution of a fluid with specified shear viscosity, in analogy to what is done within hydrodynamical simulations.

We use the CE approach to relate shear viscosity to temperature, cross section and density
which is in agreement with Green-Kubo correlator results \cite{Plumari:2012ep}.
Therefore, we fix $\eta/s$ and compute the pertinent total cross section by mean of the relation
\begin{equation}
\sigma_{tot}=\frac{1}{5}\frac{T}{\rho \, g(a)} \frac{1}{\eta/s}
~, 
\label{eq:sigmaTT}
\end{equation}
which is valid for a generic differential cross section $d\sigma/dt \sim \alpha_s^2/(t-m_D^2)^2$
as proved in~\cite{Plumari:2012ep}.
In the above equation $a=m_D/2T$, with $m_D$ the screening mass regulating the angular dependence
of the cross section, while  
\begin{eqnarray}
g(a)&=&\frac{1}{50}\! \int\!\! dyy^6
\left[ (y^2{+}\frac{1}{3})K_3(2 y){-}yK_2(2y)\right]\!
h\left(\frac{a^2}{y^2}\right)\nonumber\\
&&
\label{g_CE}
\end{eqnarray}
where $K_n$ is the Bessel function and the function $h$ relates the transport cross section to the total one $\sigma_{tr}(s)= \sigma_{tot} \, h(m_{D}^2/s)$
being  $h(\zeta)=4 \zeta ( 1 + \zeta ) \big[ (2 \zeta + 1) ln(1 + 1/\zeta) - 2 \big ]$. 
The $g(a)$ is the proper function accounting for
the correct relaxation time $\tau_{\eta}^{-1}=g(a) \sigma_{tot} \rho$ associated
to the shear viscosity transport coefficient. 
For isotropic cross section, i. e. $m_D\to \infty$, the function $g(a)$ is equal to $2/3$
and Eq.(\ref{eq:sigmaTT}) reduces to the relaxation time approximation
with $\tau^{-1}_{\eta} =\tau^{-1}_{tr} = \sigma_{tr} \rho$, while for finite value of $m_D$, which means anisotropic scatterings, $g(a)<2/3$.

We notice that, in the regime where viscous hydrodynamics applies, the specific microscopic details of the cross section are irrelevant, and our approach is an effective way to employ transport theory to simulate a fluid at 
a given $\eta/s$~\cite{Ruggieri:2013ova,Plumari:2015sia}.

\section{Flux tube decay in a static box}
In this Section we study the chromoelectric flux tube decay in a static box.
We assume the box is a cube with side of $5$ fm. Moreover we assume periodic boundary conditions
for particles propagating in the box. This case is of academic interest, nevertheless it is useful 
because it allows us to introduce concepts which will be useful when we will consider
more interesting case of a longitudinally expanding background, as well as it 
provides a further test of the numerical solution of Eq.~\eqref{Boltzmann}.

Given the symmetry of the problem, assuming at initial time a homogeneous electric field along the $z$ direction,
then the system will remain homogeneous along the whole dynamical evolution: the electric field
at later times as well as the currents and the invariant distribution functions will depend only
on time and not on space coordinates. Moreover it is easy to verify by Maxwell equations
that neither magnetic fields nor transverse components of the electric field can develop during
the evolution.
Within these assumptions the evolution equation for the classical field relevant for our problem is given by
\begin{equation}
\frac{d  E}{d t} 
=
- j(t)~.\label{eq:M1bi}
\end{equation}

In our simulations in the static box case we use Eq.~\eqref{eq:SF} to create particle
pairs from the color electric field: at each time step, the value of $E$ and of the volume
box being given, we compute the expected pair number, ${\cal N}$, integrating Eq.~\eqref{eq:SF}
over the volume box,
 then we distribute the ${\cal N}$ pairs uniformly in the box and 
with transverse momentum $p_T$ according to the distribution in Eq.~\eqref{eq:SF};
since the pairs have to pop out from the vacuum with vanishing total and longitudinal
momenta,  given $p_T$ we extract randomly the azimuthal angle $\phi$
which uniquely determines $p_y = p_T\sin\phi$ and $p_x=p_T\cos\phi$ 
of one of the particles in the pair; finally the momentum direction of the second particle 
is given by $\pi-\phi$.

\begin{figure}[t!]
\begin{center}
\includegraphics[width=7.5cm]{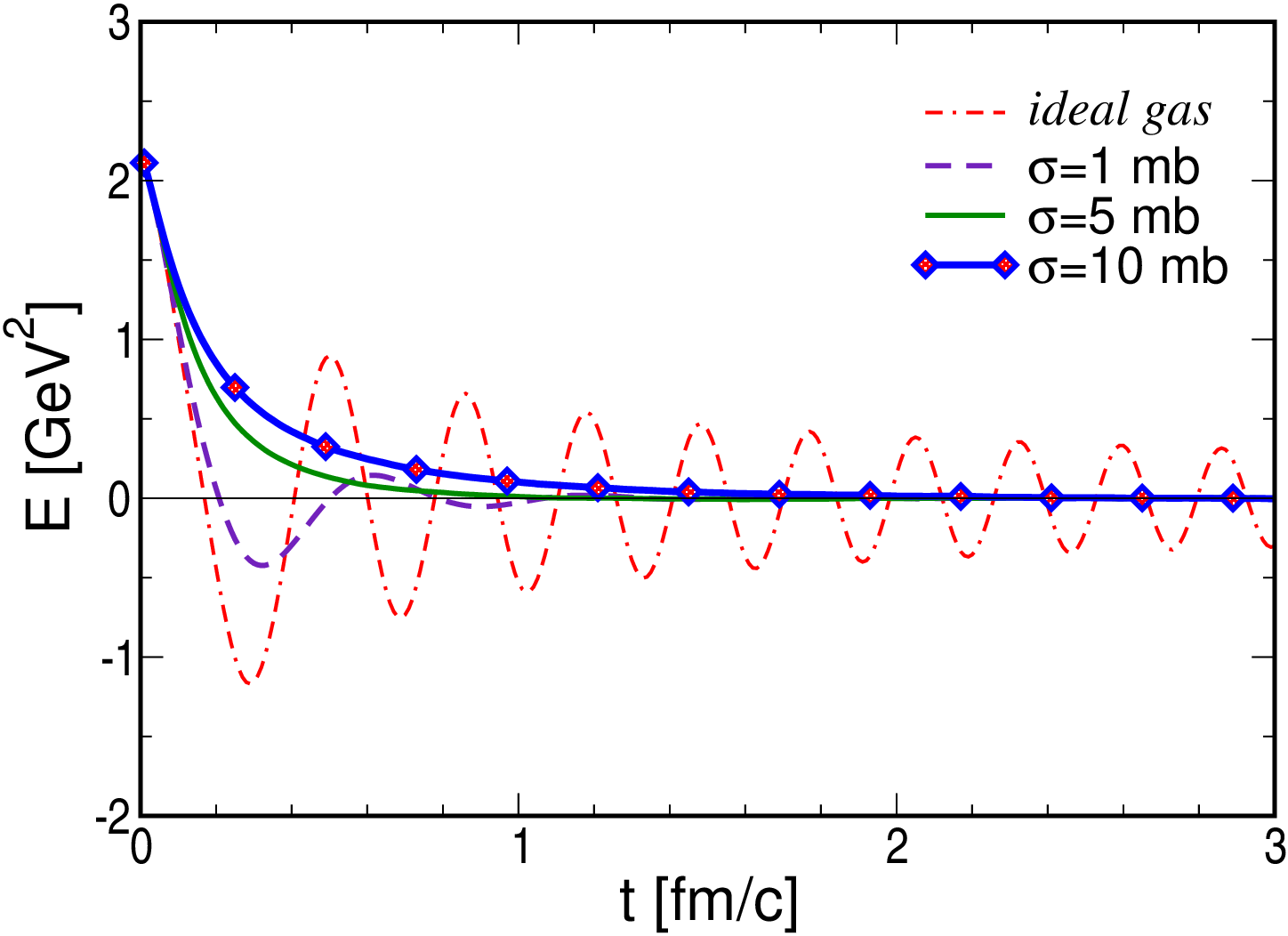}\\
\includegraphics[width=7.5cm]{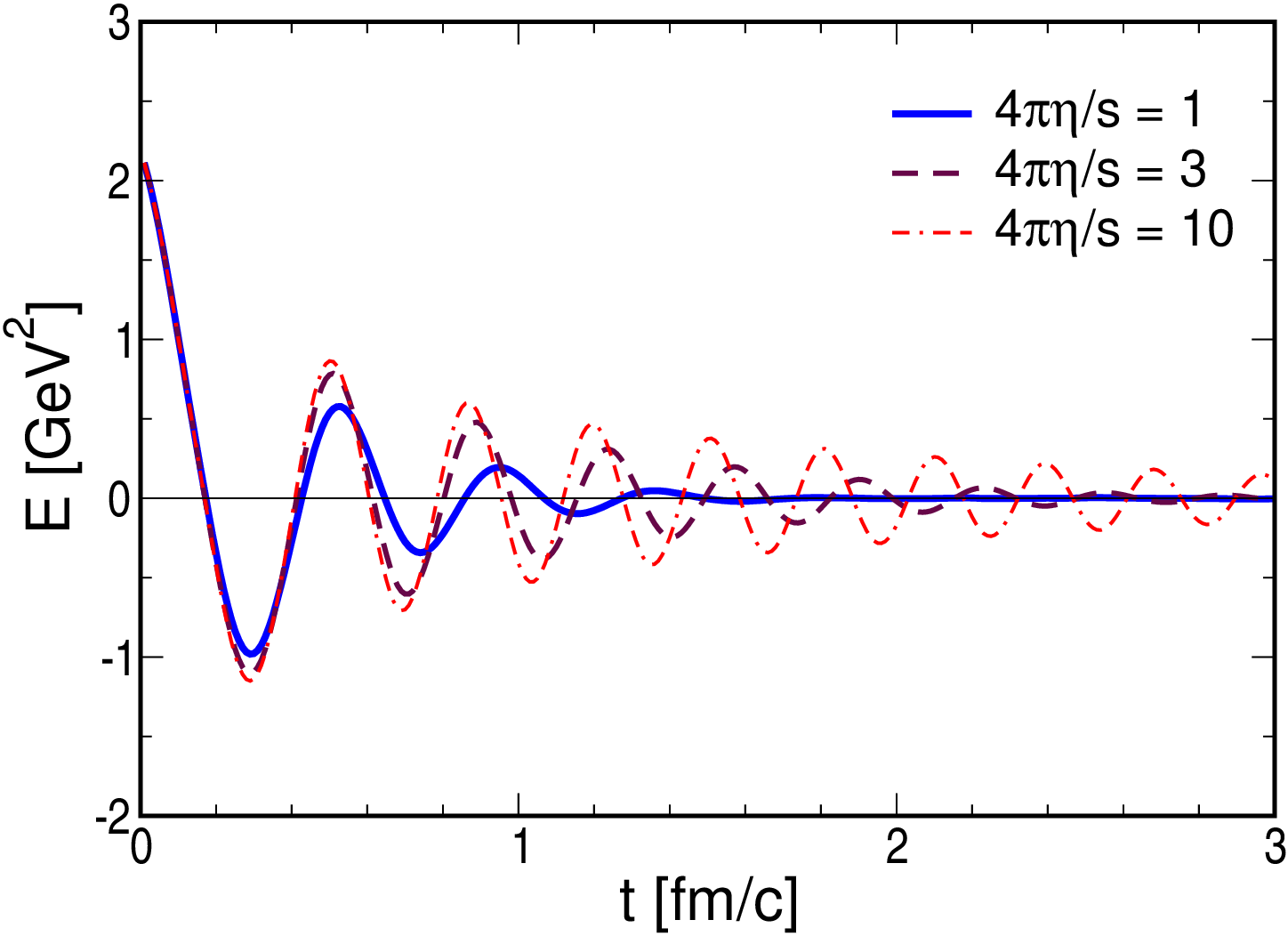}
\end{center}
\caption{\label{Fig:2} Early times evolution of the color-electric field
for the cases of calculations at fixed total cross section (upper panel)
as well as fixed $\eta/s$ (lower panel).}
\end{figure}

In Fig.~\ref{Fig:2} we plot the early times evolution of the color-electric field
for the cases of calculations at fixed total cross section (upper panel)
as well as fixed $\eta/s$ (lower panel). In the upper panel of the figure
the line labelled as ideal gas means a calculation with zero cross section.

From the results in the upper panel of Fig.~\ref{Fig:2} we notice that for small coupling 
among the particles, the electric field rapidly decays through the Schwinger mechanism
then it evolves with damped oscillations: the smaller the coupling is, the less efficient
the damping is. Eventually for strong enough coupling the oscillations disappear and the 
color-electric fields just decays according to a power law. 
This dependence on the coupling
strength is very easy to understand and is in last analysis due to the dependence of the
electric conductivity of the plasma on the particle interaction strength \cite{Puglisi:2014sha}. 
In fact at initial times the particles are produced by the Schwinger effect
with zero longitudinal momentum, then in case the coupling is small 
they are accelerated by the electric field thus generating
an electric current $j_M$ which at first gains energy from the field thus lowering its magnitude;
because of the small coupling particles's momenta is not randomized thus currents
are efficiently produced by the field.
At some point of time the field is zero but the current $j_M$ is still nonvanishing and positive
hence causing a sign flip of the electric field and a negative acceleration
of the charges, resulting eventually in  $j_M < 0$ and an increase of the electric field.
This process causes the field oscillations we observe in Fig.~\ref{Fig:2}.
 
 On the other hand, for large values of the coupling the scattering processes
 among the particles randomize momenta hence causing $j_M\approx 0$
 and the time evolution of the field is a pure decay by the Schwinger effect.
To consider this point more closely we write down the analytical form of 
Equation~\eqref{eq:jD}, namely
\begin{equation}
j_D = \frac{\zeta(5/2)}{4\pi^3E}
\left[
\frac{4-\sqrt{2}}{4}\sum_{c=1}^3 {\cal E}_{0c}^{5/2}
+
\sum_{j=1}^{N_f}\sum_{c=1}^3  {\cal E}_{jc}^{5/2}
\right]
\end{equation}
For example in the case one considers only the decay into gluons and assuming for simplicity
$\sigma_g=0$ in Eq.~\eqref{eq:sigma} one has for $j_M = 0$
the equation
 \begin{equation}
 \frac{d E }{d t} = -\sigma_D E^{3/2}~,
 \label{eq:Mwout}
 \end{equation}
where we have introduced the quantity
\begin{equation}
\sigma_D =\frac{\zeta(5/2)}{4\pi^3}
\frac{4-\sqrt{2}}{4}
\sum_{c=1}^3|g Q_{0c}|^{5/2}~,
\label{eq:sigma_D}
\end{equation}
which is related to the Schwinger effect and is independent on the coupling among the particles. 
The solution of Eq.~\eqref{eq:Mwout} can be found as
\begin{equation}
E=\frac{4E(0)}{(2 + \sigma_D \sqrt{E_0} t)^2}~,
\label{eq:ooopp}
\end{equation}
showing that in absence of a conductive current the decay of the chromoelectric field is purely
power law.

In the lower panel of Fig.~\ref{Fig:2} we plot the early times evolution of the color electric field
keeping $\eta/s$ fixed. The calculations have been performed assuming an isotropic
cross section, and the relation among the total cross section and $\eta/s$ 
is computed by the CE method \cite{Plumari:2012ep} which, 
as proved in \cite{Plumari:2012ep}, is in agreement with the
relaxation time approximation for the case of an isotropic differential cross section.
From the qualitative point of view these results do not differ from those
we obtain for calculations at fixed cross section, 
and they can be understood in the same way the results
we obtained in the case of a fixed cross section.

\begin{figure}
\begin{center}
\includegraphics[width=7.5cm]{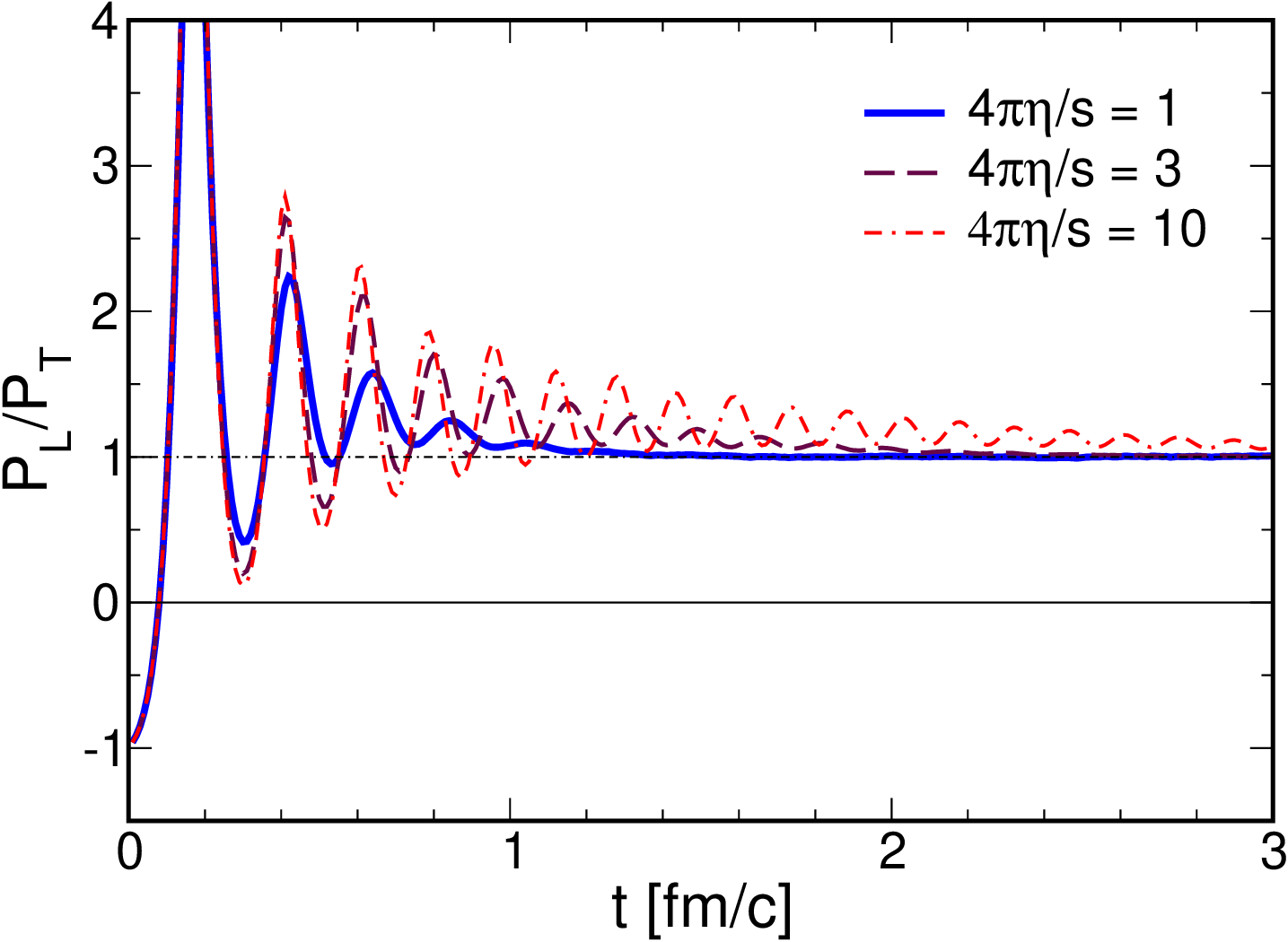}
\end{center}
\caption{\label{Fig:3} Time evolution of the ratio $P_L/P_T$
for the static box, for several values of $\eta/s$.}
\end{figure}

In Fig.~\ref{Fig:3} we plot the time evolution of the ratio $P_L/P_T$ for the case
of the static box, for several values of $\eta/s$. As expected we find that for small values
of $\eta/s$ the system is very efficient in removing the initial anisotropy.
We notice that the initial longitudinal pressure is negative and $P_L/\varepsilon=-1$;
indeed at initial time the system is made of pure color electric field
for which $T^{\mu\nu}=\text{diag}\left(\varepsilon,P_T,P_T,P_L\right)$
with $\varepsilon=E^2/2$, $P_T=\varepsilon$ and $P_L = -\varepsilon$.
On the other hand as soon as particles are produced by the Schwinger effect, they give a positive contribution 
to the longitudinal pressure and the field magnitude decreases, eventually leading to a positive
pressure. Moreover because of the strong interactions among the particles produced,
$P_L$ and $P_T$ tend to the same value within a very short time range, $\tau_{iso}\approx 1$ fm/c.

We notice that regardless of the value of $\eta/s$ the system tends to remove
the initial pressure anisotropy, which is quite natural in the case of the static box since collisions
always lead to the equilibrium state. However isotropization times
are different and indeed in the case of a weakly coupled system, see for example
the case $4\pi\eta/s=10$ in Fig.~\ref{Fig:3}, the equilibrium state is reached
in a much larger time than in the case of $4\pi\eta/s=1$.
Moreover the oscillations of the electric field strength in Fig.~\ref{Fig:2} lead 
to many oscillations of $P_L/P_T$ in the case $4\pi\eta/s=10$, such oscilations
being damped effectively within the first fm/c in the case $4\pi\eta/s=1$.

\section{Flux tube with boost-invariant longitudinal expansion}
In this Section we study the effect of a boost invariant longitudinal expansion
on pair production from the decay of a chromoelectric flux tube.
We assume the expansion takes place along the direction
of the electric field; moreover the dynamics is invariant for boosts 
along the longitudinal direction. For this case we will discuss more results
because it is closer to the description of the central rapidity region in the early
stages of a relativistic heavy ion collision.
Once again we closely follow the formalism
of~\cite{Ryblewski:2013eja}. 

We remind very quickly the relevant equations in this case. The evolution equations for the
color electric field are given by the pair of Maxwell equations
\begin{equation}
\frac{\partial E}{\partial z} = \rho~,
~~~~~\frac{\partial  E}{\partial t} = -j~,
\label{eq:Max1}
\end{equation}
where both current and charge densities are computed in the
laboratory frame. Assuming boost invariance
along the longitudinal direction implies that ${\cal E}_z$ depends only on proper time
$\tau=\sqrt{t^2 - z^2}$. We can combine the two Eqs.~\eqref{eq:Max1} to form a boost invariant
equation, namely
\begin{equation}
\tau\frac{d E}{d \tau}
= z\rho - t j~,
\end{equation}
which can be rewritten as
\begin{equation}
\frac{d E}{d \tau}
= \rho\sinh\eta - j\cosh\eta~,
\label{eq:wq}
\end{equation}
where the right hand side corresponds to the (minus) electric current computed in the reference frame
where the time is the proper time, namely the local rest frame
of the fluid. Equation~\eqref{eq:wq} is in agreement with the boost invariant form
of Maxwell equation used in \cite{Ryblewski:2013eja}, see also \cite{Gatoff:1987uf}.

To solve Eq.~\eqref{eq:wq} we adopt a finite difference scheme and 
prepare a box with a square cross section 
in the transverse direction, with $-x_{max} \leq x \leq x_{max}$
and $-y_{max} \leq y \leq y_{max}$, and with
cells in space-time rapidity, fixing the range of $\eta$ in which we distribute the produced
particles by $-\eta_{max} \leq \eta \leq \eta_{max}$ with $\eta_{max} = 2.5$.
This implementation corresponds to have a box with a longitudinal expansion since 
from the well known equations of relativistic kinematics one gets
$z_{max} = t\tanh\eta_{max}$ which, for $\eta_{max}$ sufficiently large,
corresponds to a wall moving at ultrarelativistic speed along the longitudinal direction.
We therefore distribute the pairs created by the decay of the flux tube with uniform probability
in each of the cells in $(x,y,\eta)$.

To take into account the longitudinal expansion Eq.~\eqref{eq:SF} has to be modifed as \cite{Ryblewski:2013eja}
\begin{equation}
\frac{dN}{d^4x d^2 p_T dy} = {\cal R}(p_T)
\delta(w)v~,
\label{eq:SF2}
\end{equation} 
where  $y$ denotes the momentum rapidity; 
${\cal R}(p_T)$ depends only on transverse momentum and it is not affected
formally by the expanding geometry; moreover we have introduced the two boost-invariant variables
\begin{equation}
w=tp_z - z p_0~,~~~v=p_0 t-z p_z~,
\label{eq:wv}
\end{equation}
and $\delta(w)$ affects the longitudinal momentum distribution
by forcing the condition
\begin{equation}
p_z = \frac{z}{t}p_0 = p_T \sinh\eta
\end{equation}
for the produced pairs, with $p_0=\sqrt{p_T^2 + p_z^2}$; such a condition is equivalent to assume that momentum rapidity
of the produced pair is equal to space-time rapidity. 
The procedure we implement to create pairs in the case of the boost invariant
longitudinal expansion is very similar to the one we have described in the case
of the static box, the only difference being that in the present case
we iterate the static box procedure for each rapidity cell.

\subsection{Field decay, particle production and spectra}

\begin{figure}[t!]
\begin{center}
\includegraphics[width=7.5cm]{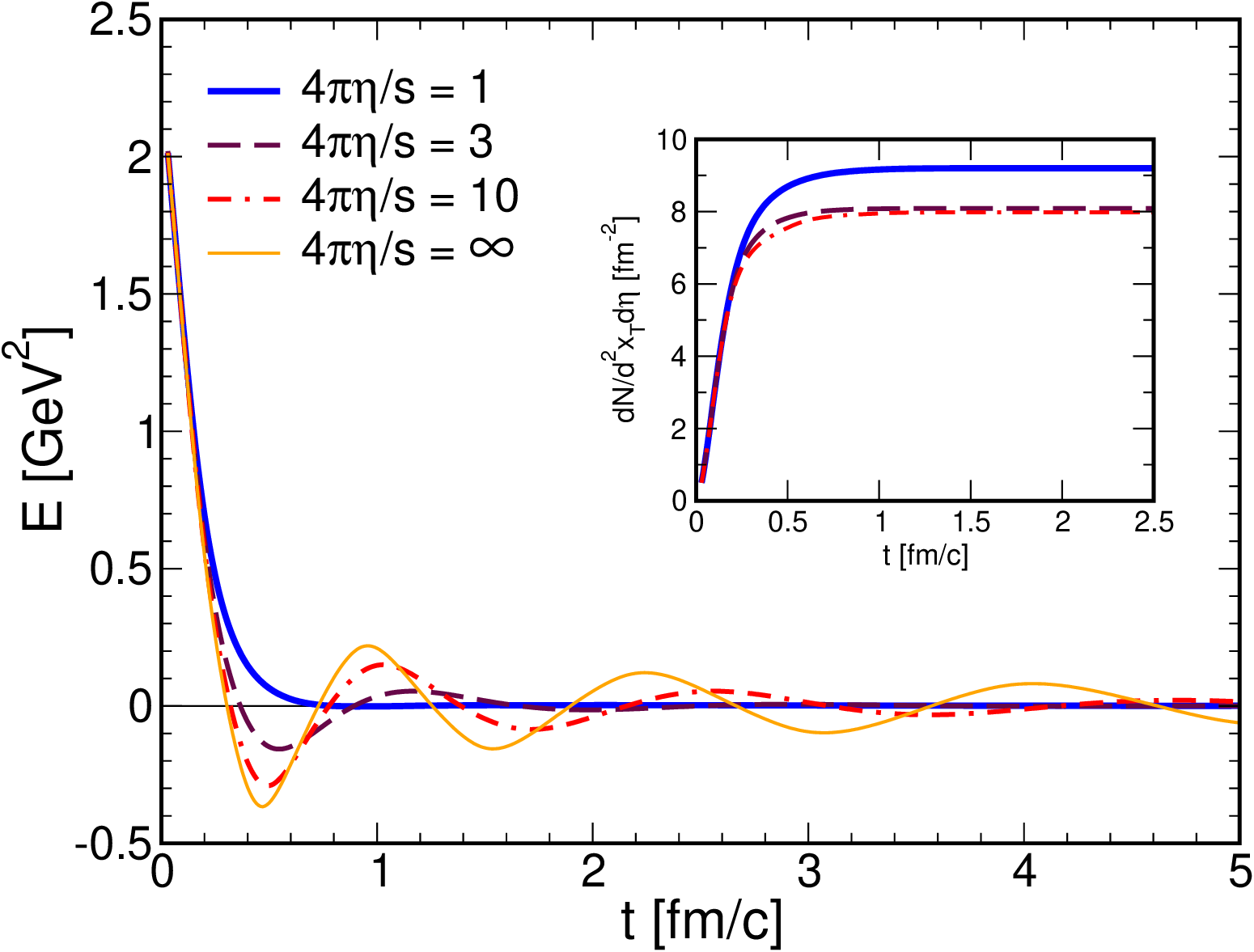}
\end{center}
\caption{\label{Fig:1a}  Chromoelectric field strength (main panel)
and particle number produced per unit of transverse area and rapidity (inset panel)
as a function of time. The electric field is averaged in the central rapidity region 
$|\eta|<0.5$.}
\end{figure}
In Fig.~\ref{Fig:1a} we plot the color electric field strength averaged 
in the central rapidity region $|\eta|<0.5$ (main panel)
and particle number produced per unit of transverse area 
and rapidity (inset panel)
as a function of time. The calculations are shown for three different values
of $\eta/s$ (calculations at fixed total cross section give similar results);
the relation among the total cross section and $\eta/s$ we have used in the simulation
is the CE relation with an isotropic cross section,
see  Eqs.~\eqref{eq:sigmaTT} and ~\eqref{g_CE}.
The electric field is averaged in the central rapidity region $|y|<0.5$ because in this region
within a $10\%$ of accuracy
one has $t\approx\tau$ with $t$ corresponding to the laboratory frame in which the system
expands, so a comparison with \cite{Ryblewski:2013eja} where the dynamics is followed
in $\tau$ is more meaningful. The initial value we have used in the simulations is
$E(t=0) = 2.2$ GeV$^2$ in agreement with the LHC 
case of \cite{Ryblewski:2013eja}  but we obtain similar results by changing this value.

As for the case of the static box we find that for the case of boost invariant longitudinal expansion
the chromoelectric field experiences a rapid decay for small values of $\eta/s$.
Once again this is due to the fact that in this case the coupling among particles is large,
meaning collisions are very effective in randomizing particle momenta in each cell hence
damping conductive currents that might sustain the field. On the other hand for intermediate and
large values of $\eta/s$ the electric field experiences stronger fluctuations during time evolution.

In the inset of Fig.~\ref{Fig:1a}  we plot the number of produced gluons per unit of transverse area
and rapidity versus time. 
We find that regardless of the value of $\eta/s$ we use in the simulation,
the particles are produced at very early times, approximately within 0.5 fm/c,
with the only expection of very few particles produced at later times
in the case $4\pi\eta/s=10$. We have checked that changing the initial value $E_0$ of the
electric field does not modify the production time in a considerable way unless 
$E_0$ is very small, namely $E_0\ll 1$ GeV$^2$.
Moreover the value of $\eta/s$ affects the conversion of the initial classical field 
to gluons only within a few percent: for example comparing the results 
for $4\pi\eta/s=1$ and  $4\pi\eta/s=3$ we find in the latter case a lowering
of less than $10\%$ on the number of particles produced.

\begin{figure}[t!]
\begin{center}
\includegraphics[width=7.5cm]{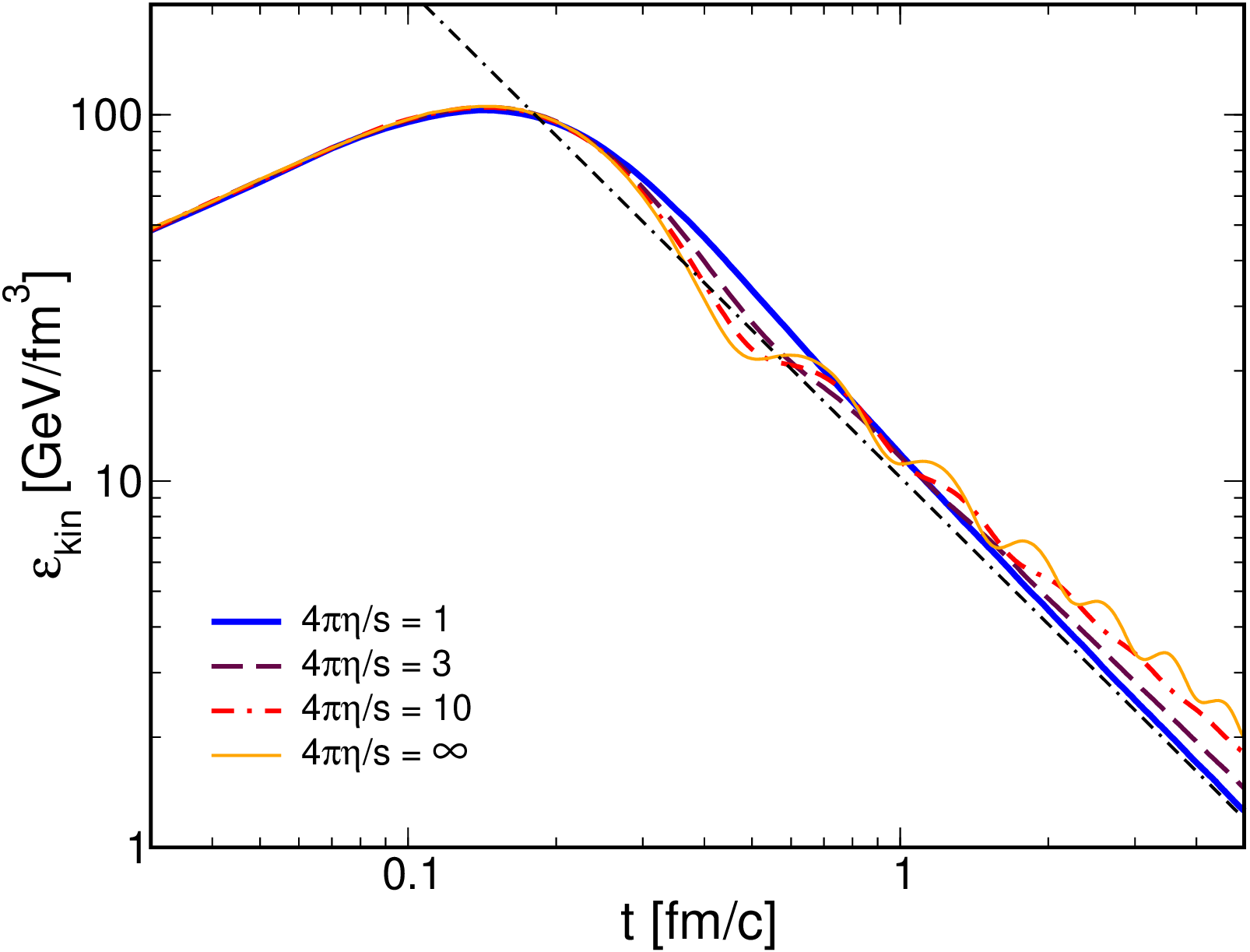}\\
\includegraphics[width=7.5cm]{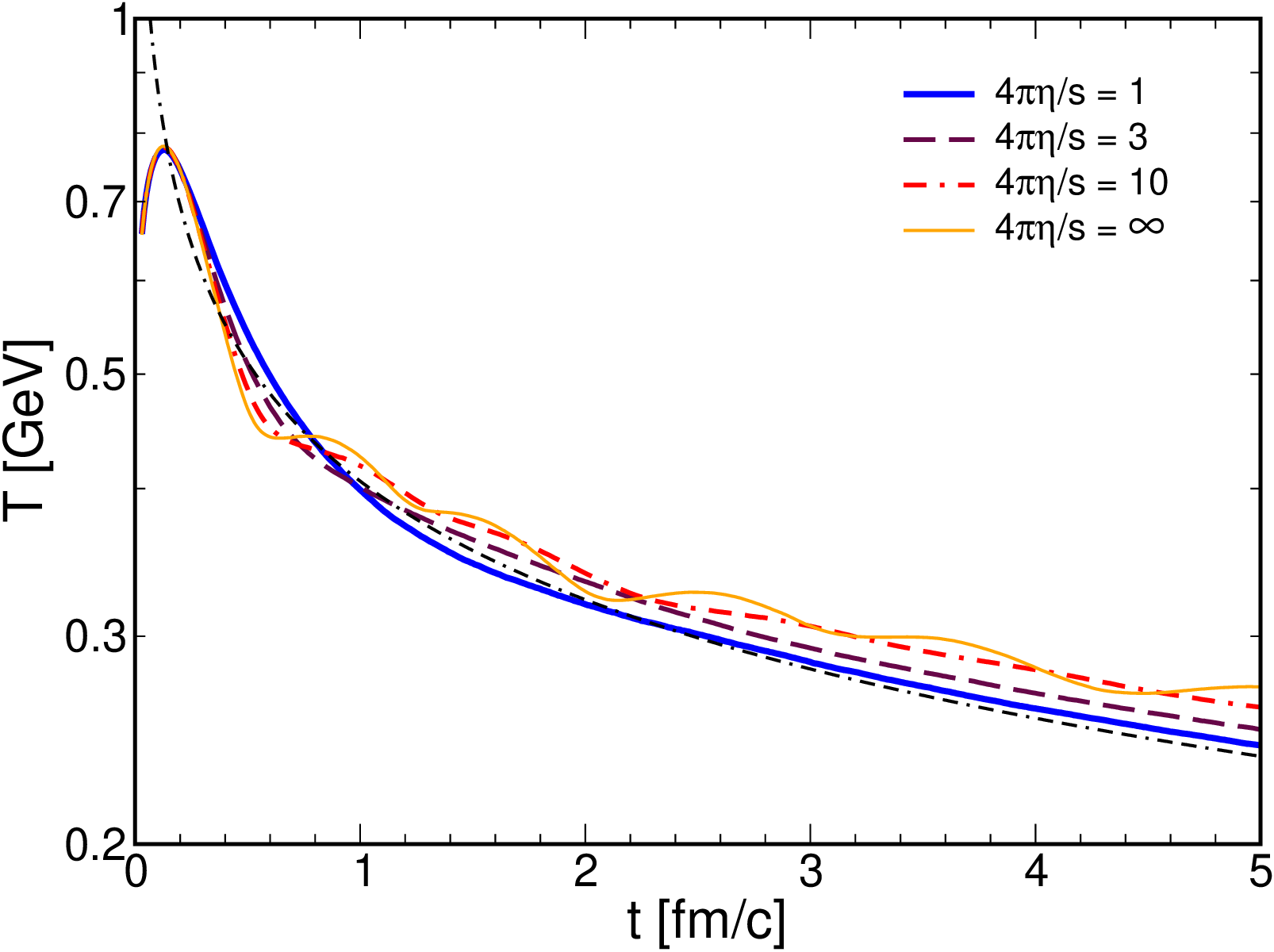}
\end{center}
\caption{\label{Fig:2aNEW}  Proper kinetic energy density (upper panel) and
local temperature (lower panel). All the quantities are averaged in the central rapidity region 
$|\eta|<0.5$ and viscosity has been fixed using an isotropic cross section.}
\end{figure}
In the upper panel of Fig.~\ref{Fig:2aNEW} we plot the proper kinetic energy density,
 $\varepsilon_{kin}$, at central rapidity ($|\eta|<0.5$),
versus laboratory time for three different values of $\eta/s$
 (calculations at fixed total cross section give similar results).
 The relation among $\eta/s$ and cross section is fixed by the CE relation
 with an isotropic cross section.
We find that in the case of small $\eta/s$, which corresponds to the case of a strongly
coupled system, the energy density decays asymptotically as $\varepsilon_{kin}\propto t^{-4/3}$
which is what is expected in the ideal hydrodynamic limit in the case of a one-dimensional
expansion, in agreement with  \cite{Gatoff:1987uf}.
For the cases of larger $\eta/s$ we find that a power law decay 
with a superimposed oscillation pattern is present.
The thin dashed line in the figure corresponds
to $t^{-4/3}$. 
These results are in good agreement with those of \cite{Ryblewski:2013eja}.

In the lower panel of Fig.~\ref{Fig:2aNEW} we plot the plasma temperature as a function of time;
it is obtained by data shown in the upper panel of  Fig.~\ref{Fig:2aNEW}  by assuming 
a perfect gas equation of state which gives $T\propto \varepsilon_{kin}^{1/4}$ with proportionality
constant being inversely proportional to the number of active degrees of freedom in the plasma. 
Our temperature is somehow
larger than the one quoted in \cite{Ryblewski:2013eja} because in the latter study both quarks
and gluons have been considered in the plasma, while in our case we only include gluons.
The thin dashed line corresponds to $t^{-1/3}$ which is the power law decay expected in the case
of a one dimensional expansion of a non viscous fluid.

\begin{figure}[t!]
\begin{center}
\includegraphics[width=7.5cm]{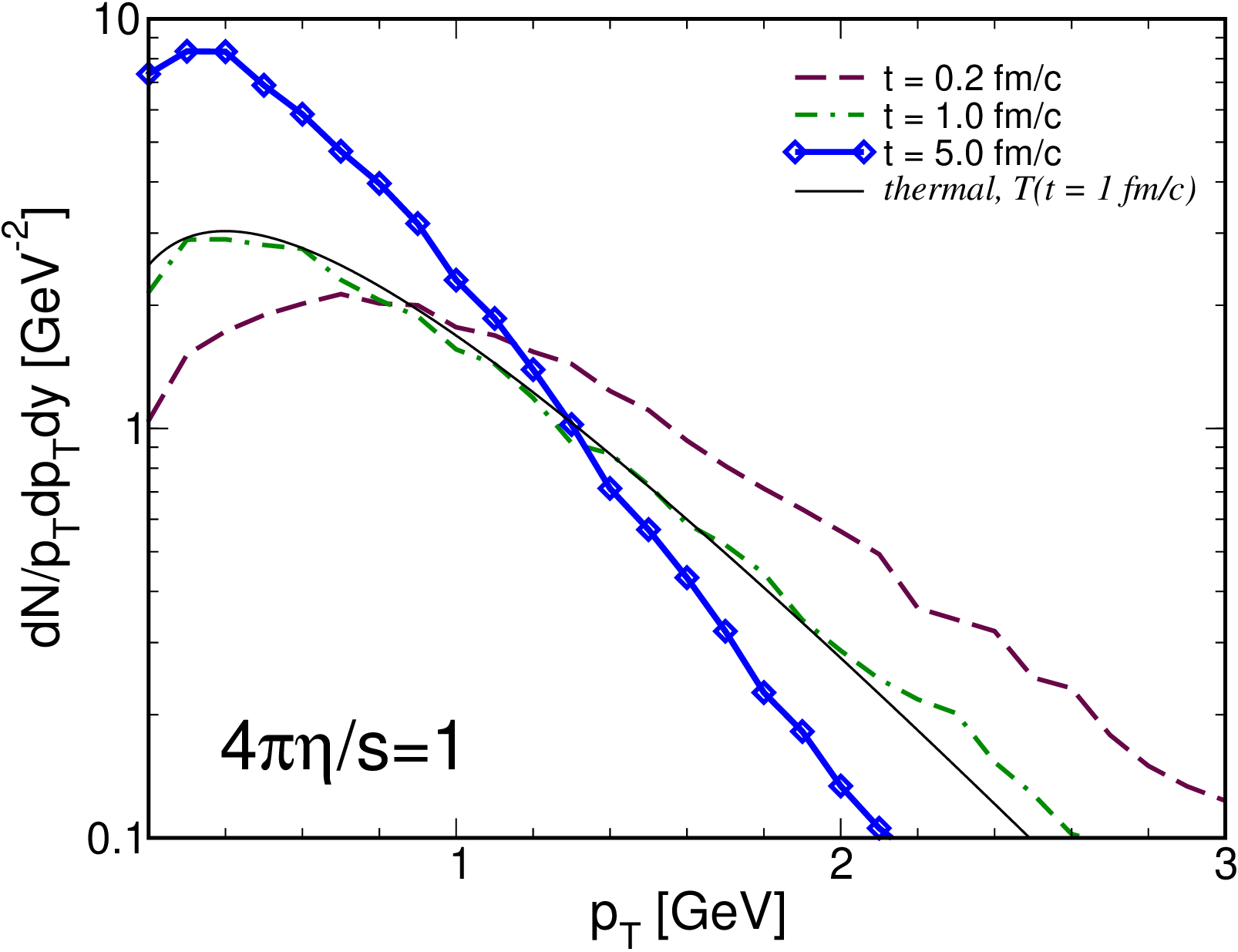}\\
\includegraphics[width=7.5cm]{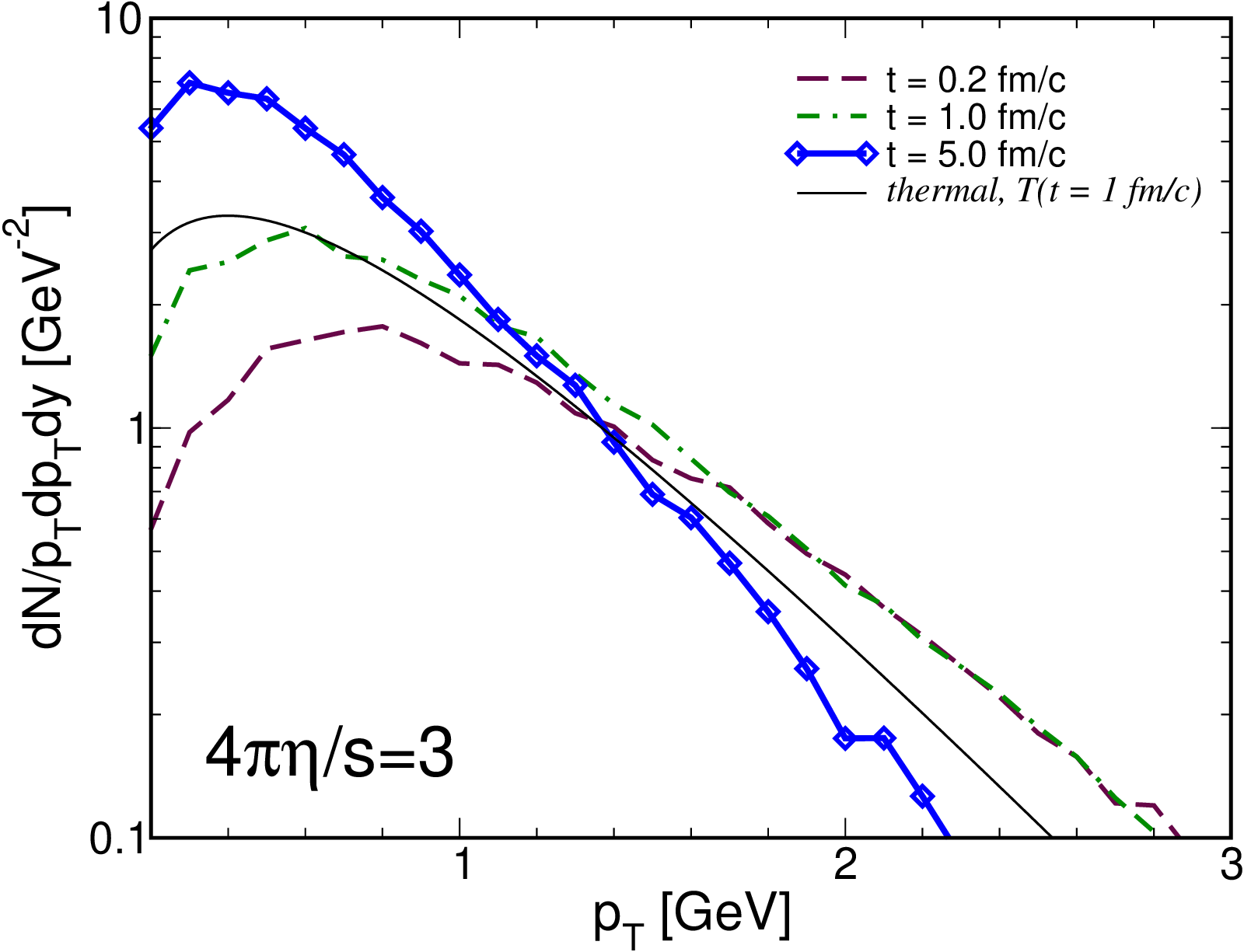}\\
\includegraphics[width=7.5cm]{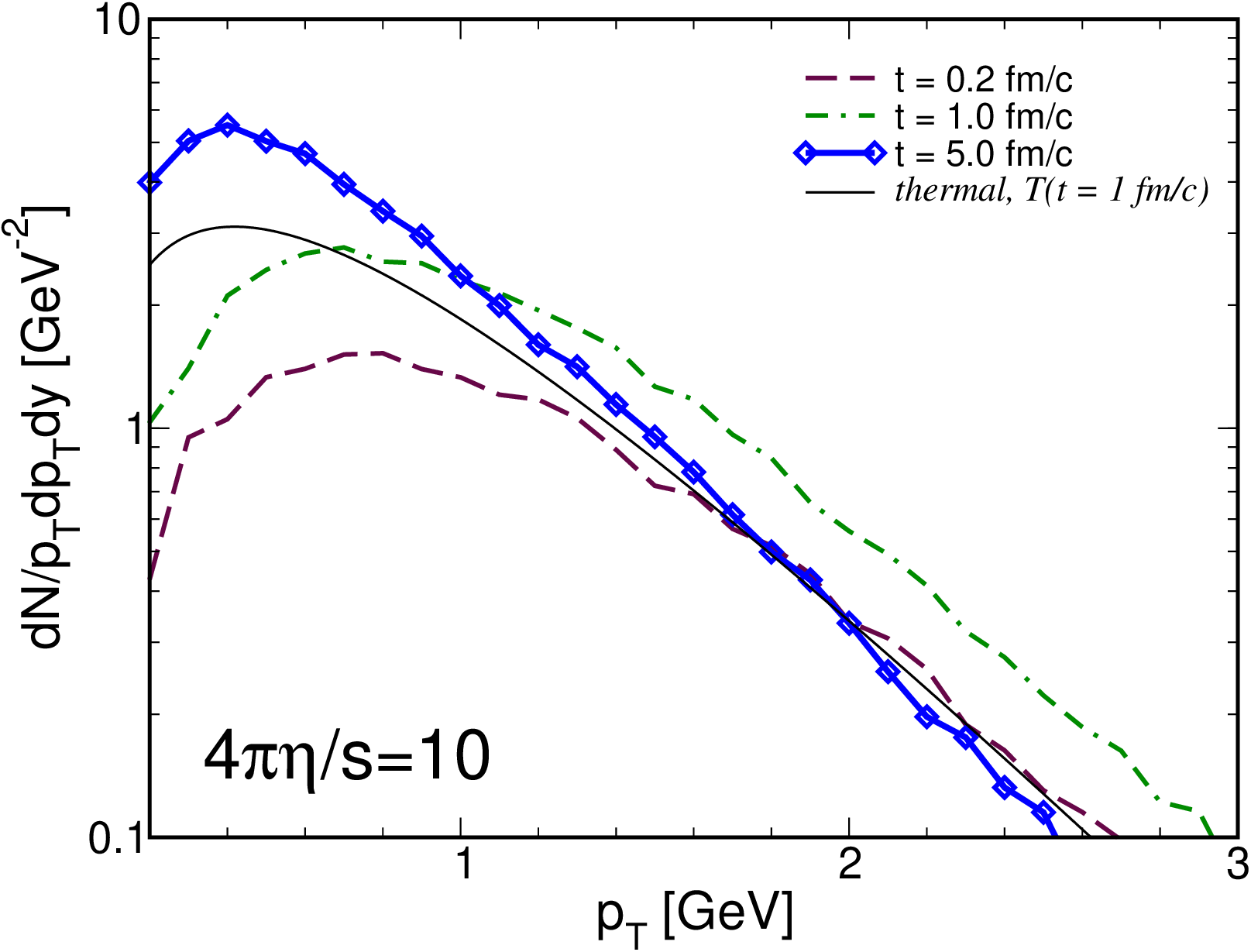}
\end{center}
\caption{\label{Fig:1ab}  Gluon spectra at midrapidity $|y|<0.5$,
for three different values of $\eta/s$. Upper panel corresponds to  $4\pi\eta/s=1$, middle panel to  $4\pi\eta/s=3$
and lower panel to  $4\pi\eta/s=10$.
For each value of $\eta/s$
the spectrum at three different times is shown. Black thin solid line 
corresponds to the thermal spectrum in Eq.~\eqref{eq:thermalspe}
at $t=1$ fm/c with temperature from data in Fig.~\ref{Fig:2aNEW}.}
\end{figure}
In Fig.~\ref{Fig:1ab} we plot the gluon spectra at midrapidity $|y|<0.5$,
for three different values of $\eta/s$. 
Upper panel corresponds to  $4\pi\eta/s=1$, middle panel to  $4\pi\eta/s=3$
and lower panel to  $4\pi\eta/s=10$.
For each value of $\eta/s$
the spectrum at three different times is shown.
The thin solid black line corresponds to a thermal spectrum, namely
\begin{equation}
\frac{dN}{p_T dp_T dy}\propto p_T e^{-\beta p_T}~;
\label{eq:thermalspe}
\end{equation}
the above relation describes a thermalized system in three spatial dimensions at the temperature $T=1/\beta$.
In the figure the thermal spectrum is computed by taking the temperature at $t=1$ fm/c
from data plot in Fig.~\ref{Fig:2aNEW} .
We find that for $4\pi\eta/s=1$ the system efficiently thermalizes:
in fact the spectrum at $t=1$ fm/c is of the form \eqref{eq:thermalspe}
with temperature given (within a $2\%$) by the result in Fig.~\ref{Fig:2aNEW},
as it is evident by comparing the thermal spectrum (black line) with simulation data 
(dot-dashed thin green line).
For $4\pi\eta/s=3$  and $4\pi\eta/s=10$ the spectrum we obtain in the numerical calculation is in some
disagreement with the thermal spectrum at the temperature plot in Fig.~\ref{Fig:2aNEW}
at $t=1$ fm/c, meaning the system is not completely thermalized in three dimensions.
Moreover the very mild 
change in the slope of the spectrum we measure from $t=1$ fm/c to $t=5$ fm/c shows that
the system does not cool down efficiently in this case, as it is expected because
the large viscosity implies that a large part of energy dissipates into heat and the system
cools down more slowly than the case of small viscosity.  

\subsection{Pressure isotropization}
\begin{figure}[t!]
\begin{center}
\includegraphics[width=7.5cm]{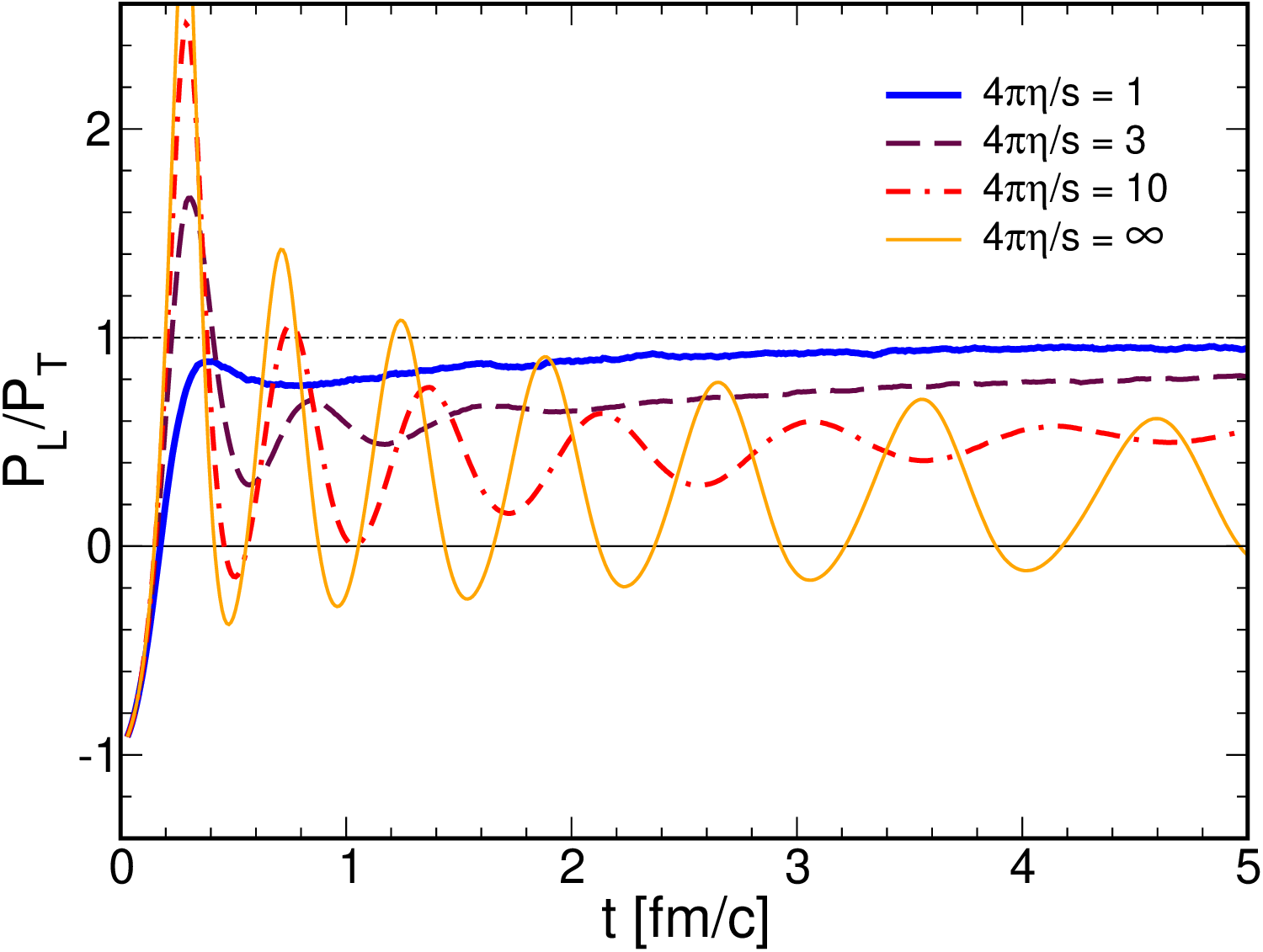}
\end{center}
\caption{\label{Fig:2a}  Proper kinetic energy density (upper panel),
local temperature (middle panel) and the ratio $P_L/P_T$ (lower panel)
as a function of time. All the quantities are averaged in the central rapidity region 
$|\eta|<0.5$ and viscosity has been fixed using an isotropic cross section.}
\end{figure}
In this subsection we discuss pressure isotropization in the case with a longitudinal expanding geometry.
In Fig.~\ref{Fig:2a} we plot the ratio $P_L/P_T$
where once again $P_L$ and $P_T$ correspond to the longitudinal and transverse
pressure respectively.  
These quantities are computed cell by cell in the local rest frame of the fluid, then
averaged in the rapidity range $|\eta|<0.5$.
The initial longitudinal pressure at initial time is negative and $P_L/P_T=-1$
because at initial time the system is made of pure longitudinal chromoelectric field.
On the other hand as soon as particles are produced, they give a positive contribution 
to the longitudinal pressure and the field magnitude decreases, eventually leading to a positive
pressure. For all the value of $\eta/s$ we consider in our simulations
we find that the time needed to the total longitudinal pressure
to be positive is about $0.2$ fm/c.
Moreover in the case $4\pi\eta/s=1$ the strong interactions among the particles 
remove the initial pressure anisotropy quite efficiently and quickly: 
in this case $P_L/P_T=0.7$ within $0.6$ fm/c, then the ratio tends
to increase towards 1 within the time evolution of the system.
This would justify the use of viscous hydrodynamics with $\tau_0\approx 0.6$ fm/c
as commonly done.

On the other hand  
the larger the $\eta/s$ of the fluid the larger the oscillations of $P_L/P_T$, compare for example
the cases $4\pi\eta/s=1$ and $4\pi\eta/s=10$ in Fig.~\ref{Fig:2a}: in the latter case
$P_L/P_T$ experiences several oscillations which follow the alternation of 
maxima of $|E|$ (corresponding to minima of $P_L$ since the field
gives a negative contribution to $P_L$) and zeros of $E$
(corresponding to maxima of $P_L$); also, at large times $P_L/P_T$
is quite smaller than 1. For intermediate values of
$\eta/s$ we still find some oscillation which become less important for smaller viscosities.
On the same footing the asymptotic value of $P_L/P_T$ approaches $1$ when the viscosity
becomes small. 

\begin{figure}[t!]
\begin{center}
\includegraphics[width=7.5cm]{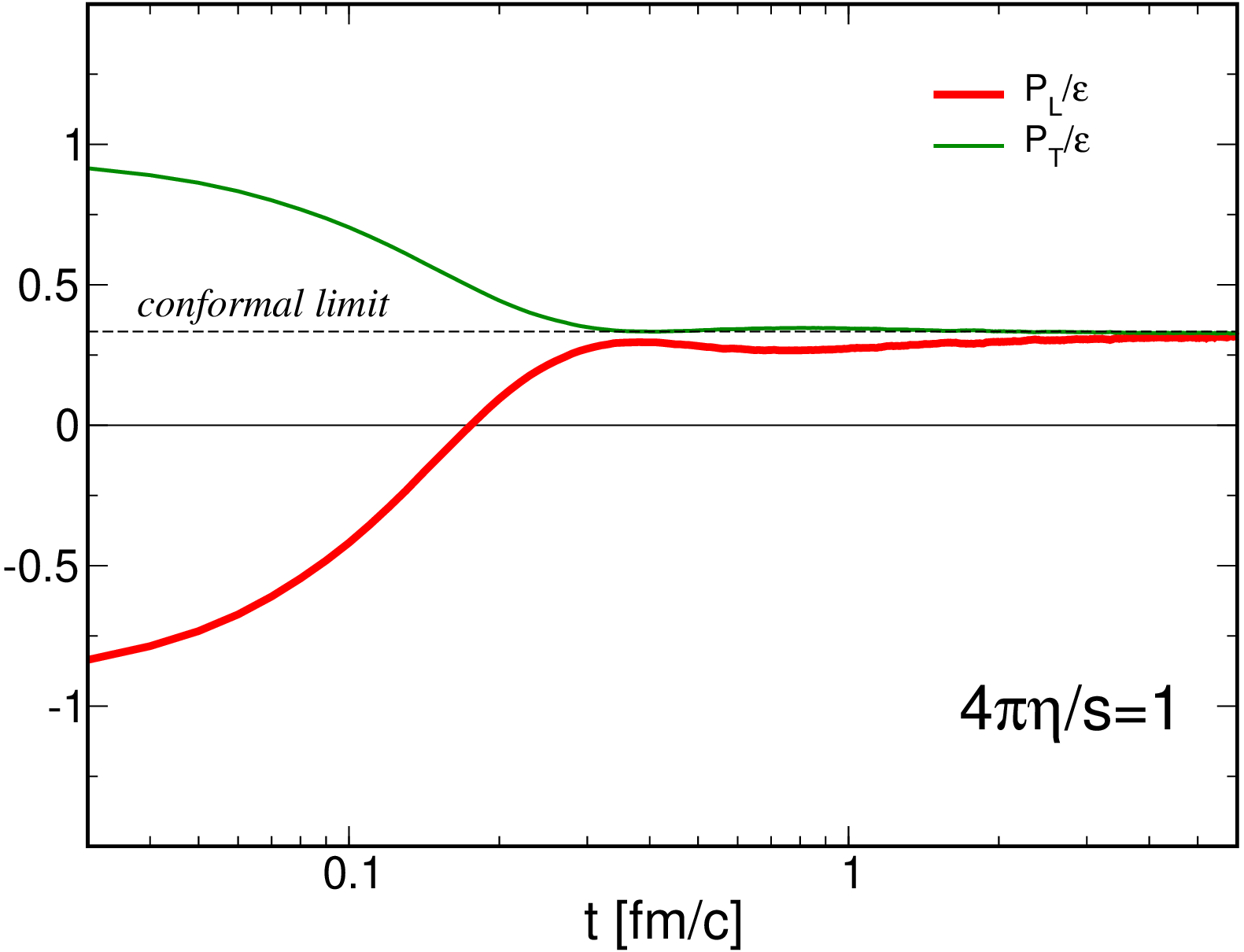}\\
\includegraphics[width=7.5cm]{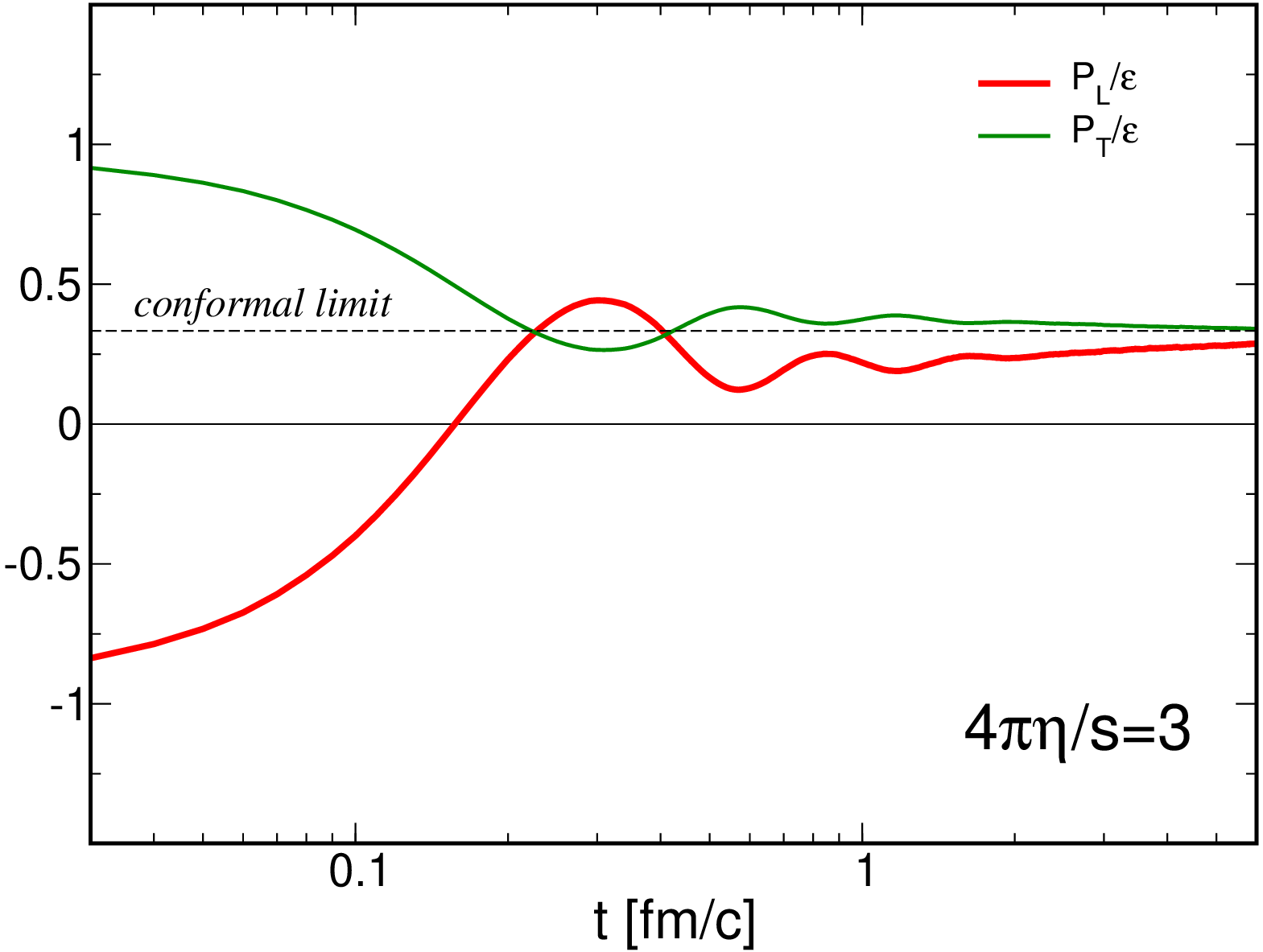}\\
\includegraphics[width=7.5cm]{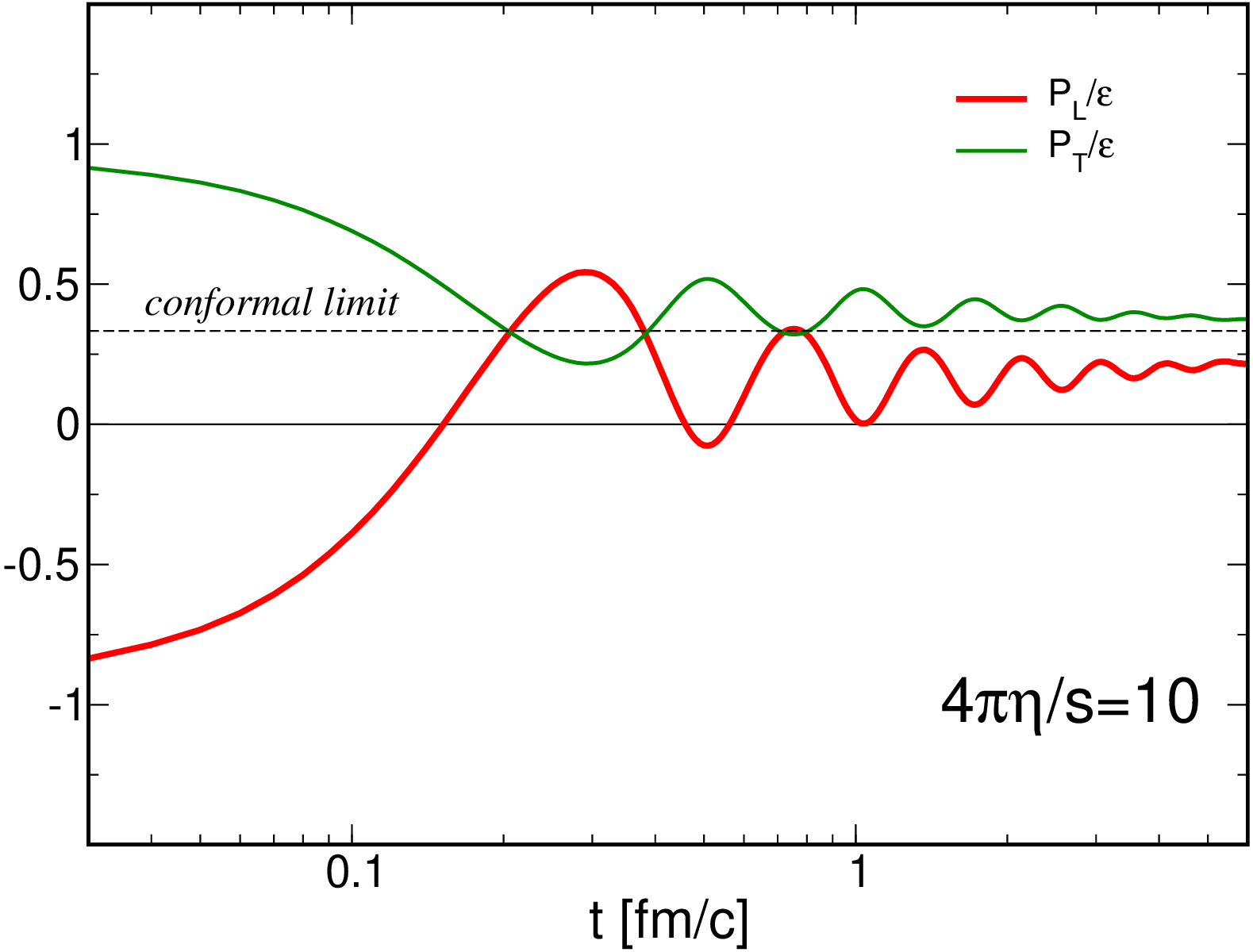}
\end{center}
\caption{\label{Fig:2aPL}  Ratios $P_L/\varepsilon$, $P_T/\varepsilon$ 
against time for $4\pi\eta/s=1$ (upper panel),
$4\pi\eta/s=3$ (middle panel) and $4\pi\eta/s=10$ (lower panel).
All the quantities are averaged in the central rapidity region 
$|\eta|<0.5$ and viscosity has been fixed using an isotropic cross section.}
\end{figure}

In Fig.~\ref{Fig:2aPL} we plot the ratios $P_L/\varepsilon$ and $P_T/\varepsilon$
as a function of time. Here $\varepsilon$ corresponds to the total energy density,
which takes into account energy density of both particles
and field. We have shown results obtained for 
$4\pi\eta/s=1$ (upper panel),
$4\pi\eta/s=3$ (middle panel) and $4\pi\eta/s=10$ (lower panel).
In Fig.~\ref{Fig:2aPL} the red solid line corresponds to $P_L/\varepsilon$
and the thin green line to $P_T/\varepsilon$. 
The thin black dashed line corresponds to the conformal isotropic limit $\varepsilon= 3P$.
One can compare the results of Fig.~\ref{Fig:2aPL} with those of \cite{Gelis:2013rba},
where a classical Yang-Mills simulation with a $3+1$D expanding geometry
is considered.
The spirit of our comparison with \cite{Gelis:2013rba} is that, given we try to solve the same physical problem
and we compute the same physical quantities, a comparison of the final results obtained within the
two approaches is interesting even if the theoretical frameworks are different. 
In the weakest coupling case considered in \cite{Gelis:2013rba}, namely $g=0.1$,
$P_L$ asymptotically relaxes towards zero, which we might obtain if we introduce
a larger viscosity than the one we consider in the present study. 
On the other hand the case $g=0.5$ \cite{Gelis:2013rba}
produces $P_L/\varepsilon$ and $P_T/\varepsilon$ which lie in between
our results for $4\pi\eta/s=3$ and $4\pi\eta/s=10$.

\subsection{Anisotropic cross sections}

\begin{figure}[t!]
\begin{center}
\includegraphics[width=7.5cm]{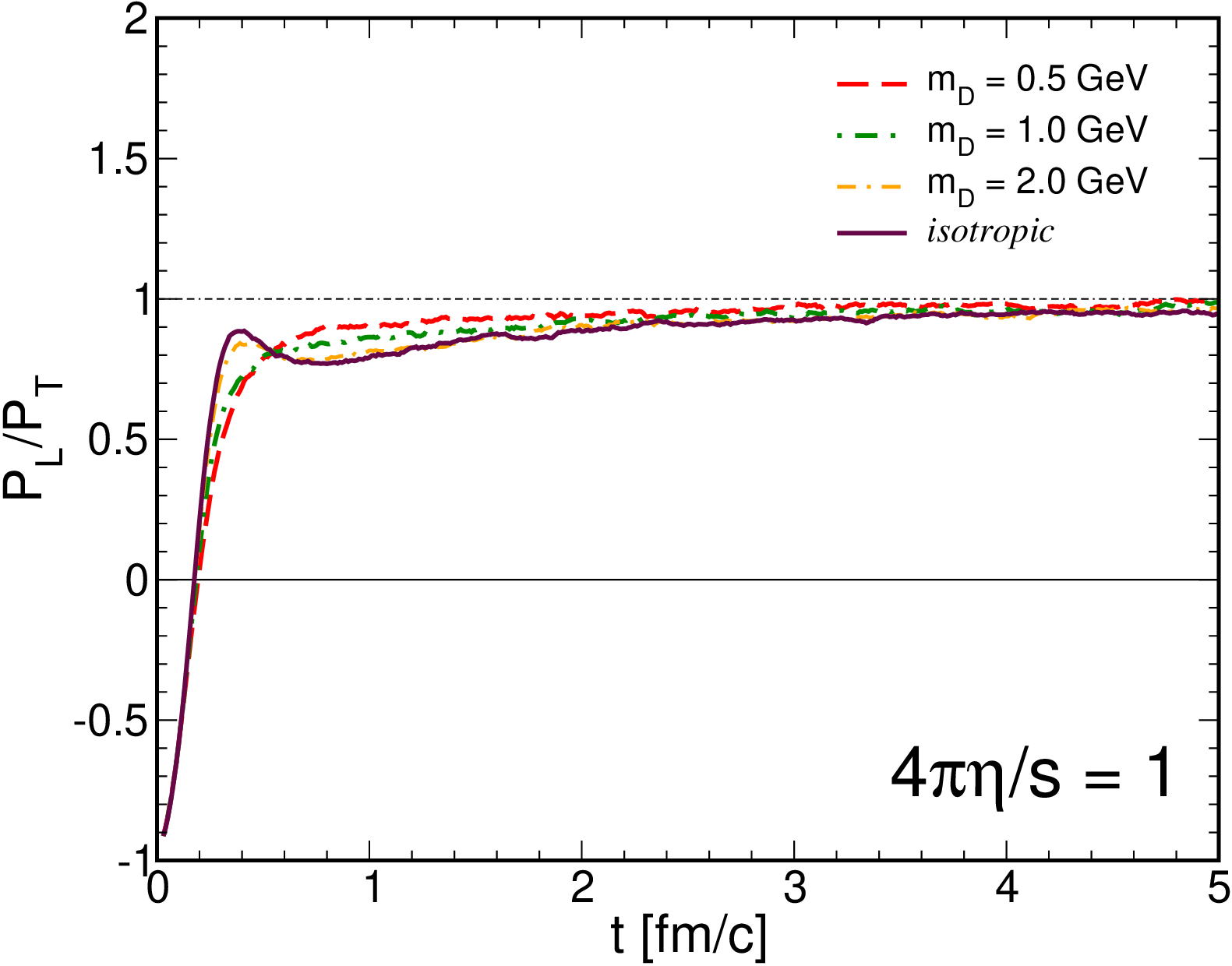}\\
\includegraphics[width=7.5cm]{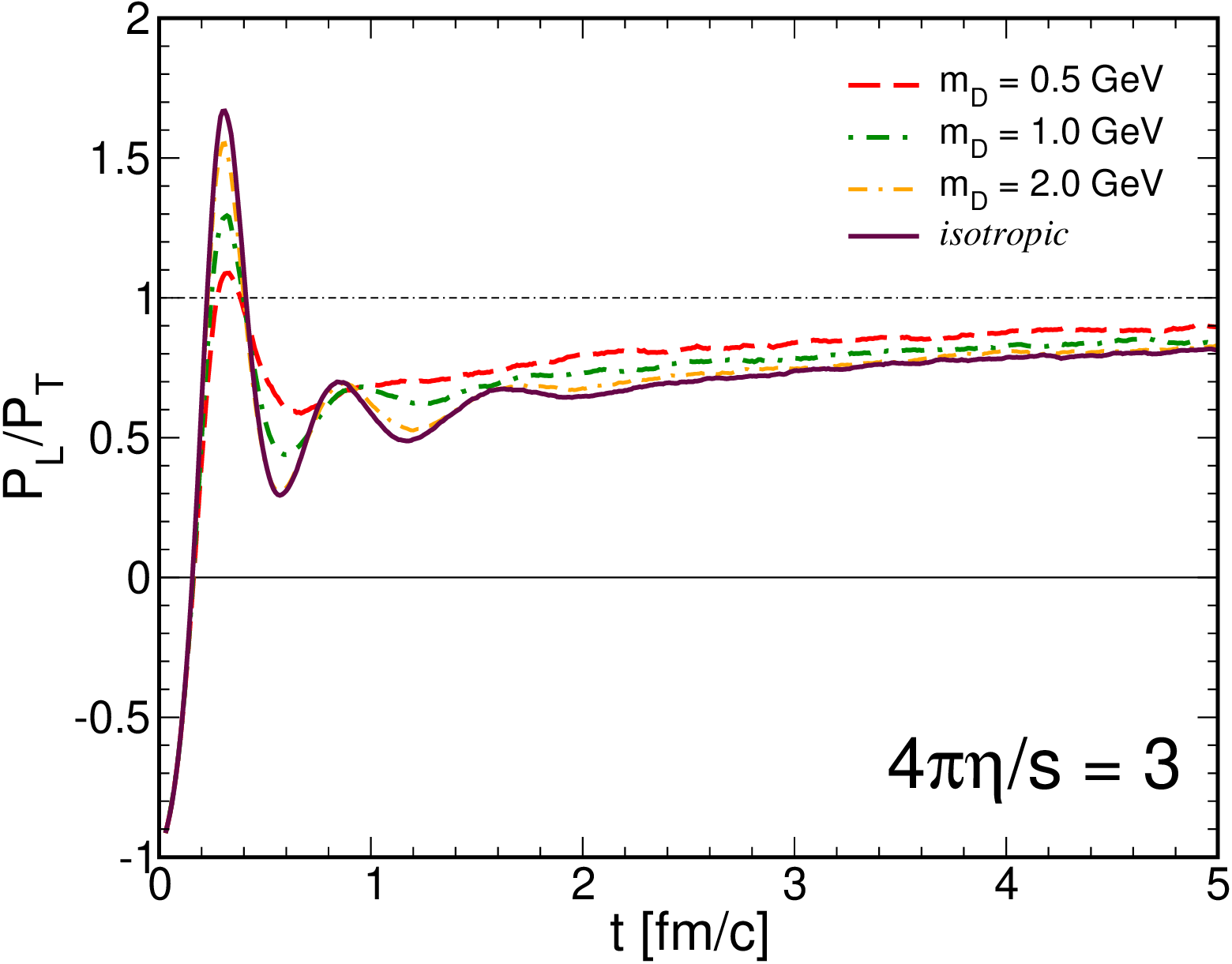}\\
\includegraphics[width=7.5cm]{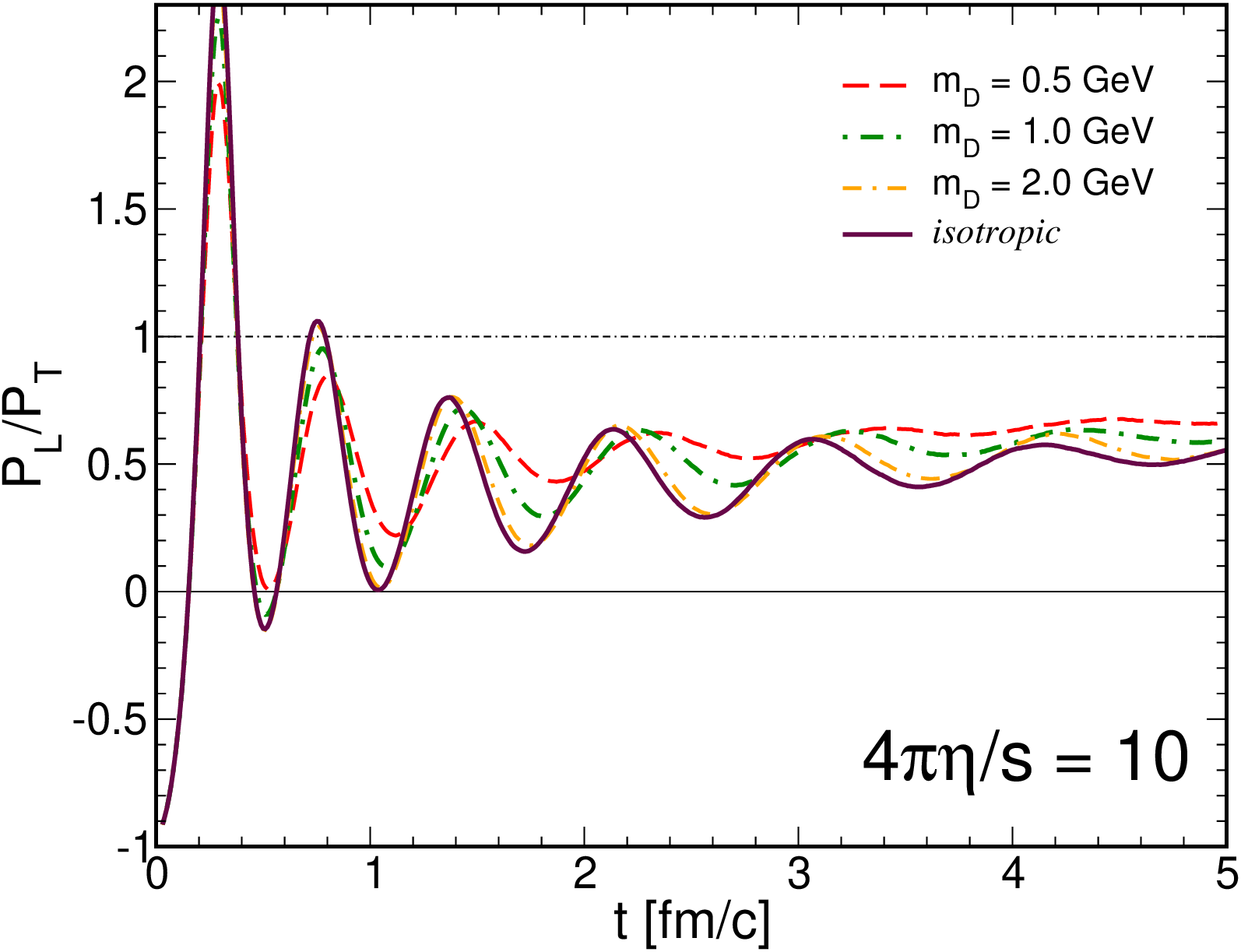}
\end{center}
\caption{\label{Fig:3ab}  Ratio $P_L/P_T$ versus time for several values of $m_D$ and $\eta/s$. 
All the quantities are averaged in the central rapidity region 
$|\eta|<0.5$. Upper panel corresponds to $4\pi\eta/s=1$,
middle panel corresponds to $4\pi\eta/s=3$ and lower panel to $4\pi\eta/s=10$.}
\end{figure}
In this subsection we study the effect of changing the microscopic cross section
from isotropic to anisotropic. In our collision integral we achieve this by tuning
the Debye screening mass in the two body cross section, leaving the value of $\eta/s$ fixed.
In Fig.~\ref{Fig:3ab} we plot the time evolution of $P_L/P_T$ for four different values of $m_D$
for the case of $4\pi\eta/s=1$ (upper panel), $4\pi\eta/s=3$ (middle panel)
and $4\pi\eta/s=10$ (lower panel), and four different
values of the Debye screening mass $m_D$. We remind that in our calculation the Debye mass
is used as an infrared regulator of the differential cross section, and as a parameter which
controls the anisotropy of the cross section: for very large values of $m_D$ the differential
cross section is isotropic, while for small values of $m_D$ we get a forward peaked cross section.

We find that lowering $m_D$ for a given value of $\eta/s$ 
the plasma oscillations tend to be damped. This can be understood
because according to~\cite{Plumari:2012ep} lowering $m_D$ while keeping
fixed $\eta/s$ amounts to increase isotropization of the distribution function; as a consequence
the conductive currents, which would sustain plasma oscillations in the late times evolution of the plasma,
are damped.
Nevertheless the effect on late time evolution of $P_L/P_T$ is quite mild.

\subsection{Chapman-Enskog versus Relaxation Time Approximation}
In this subsection we study the effect of shifting from the CE to RTA
when we relate $\eta/s$ to the total cross section. It has to be noticed that
our RTA does not correspond to the RTA used in \cite{Ryblewski:2013eja}:
in fact in \cite{Ryblewski:2013eja} RTA corresponds to an ansatz for the
collision integral in the Boltzmann equation; on the other hand in our calculations,
where we always solve the full collision integral in the Boltzmann equation,
RTA refers only to an analytical equation which connects shear viscosity to
microscopic cross section \cite{Plumari:2012ep}. We limit ourselves to 
a particular value of $\eta/s$, namely $4\pi\eta/s=3$ which is the intermediate
case we have considered in the previous subsections.

\begin{figure}[t!]
\begin{center}
\includegraphics[width=7.5cm]{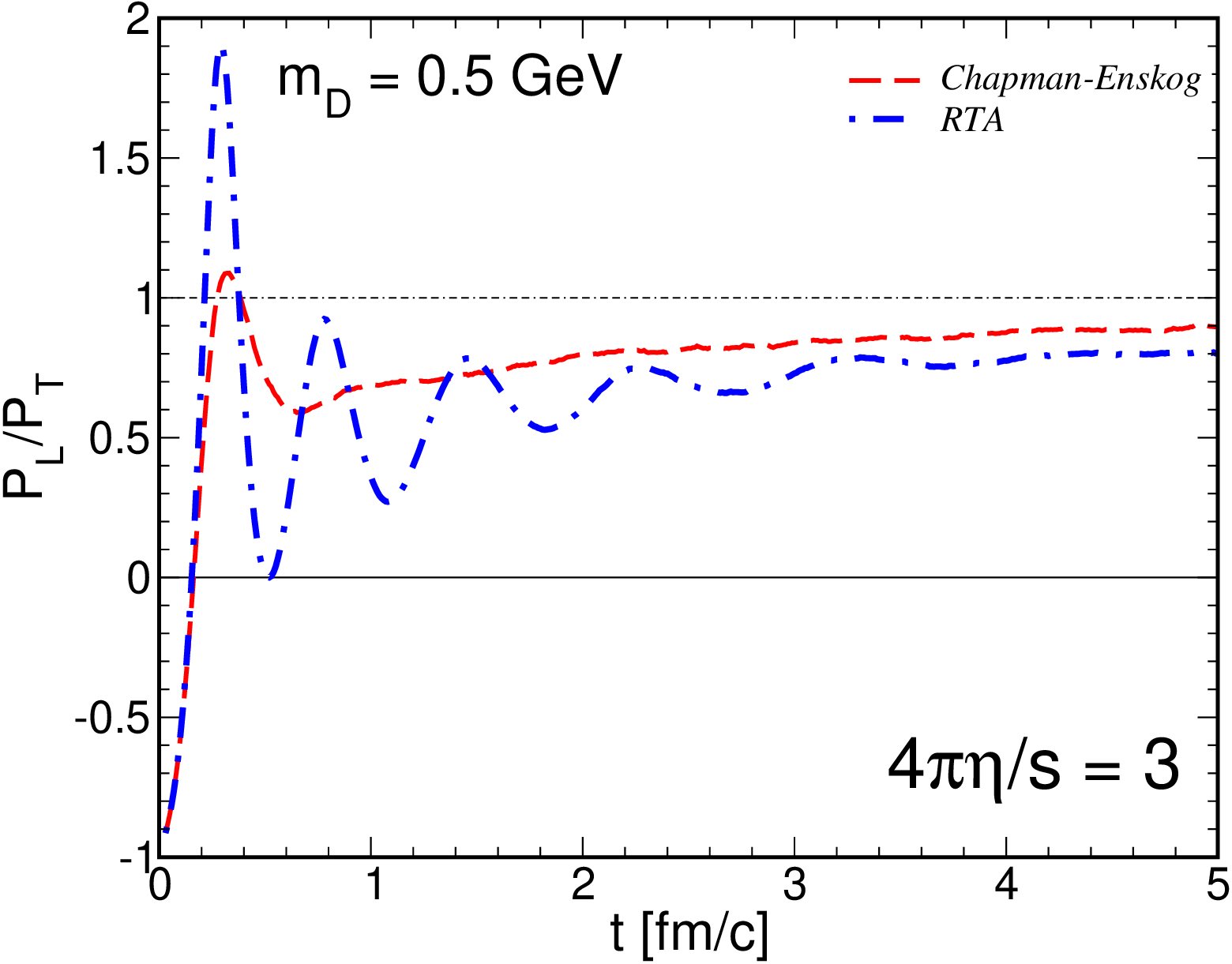}\\
\includegraphics[width=7.5cm]{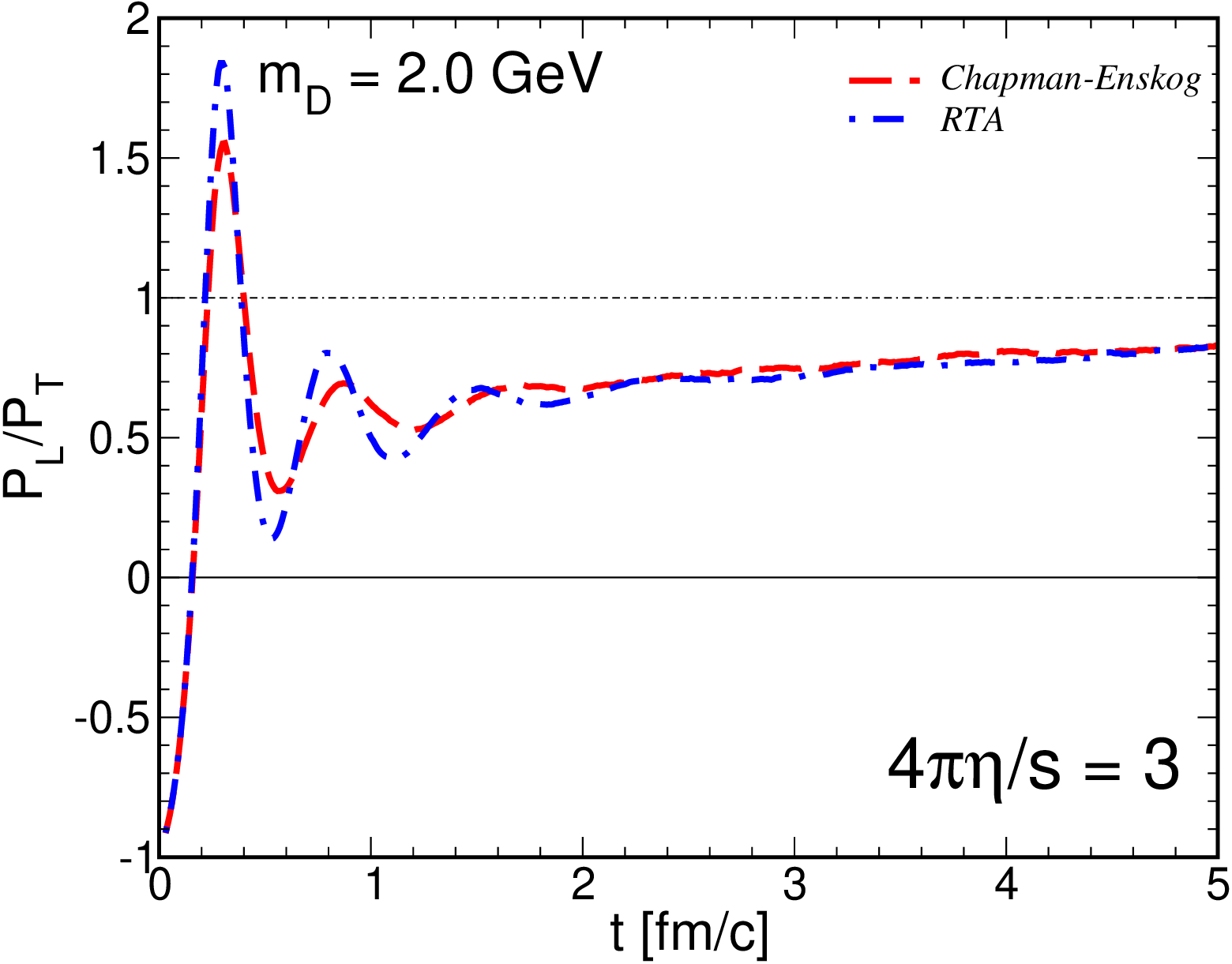}
\end{center}
\caption{\label{Fig:5abMD}  Ratio $P_L/P_T$
versus time for $4\pi\eta/s=3$ and for 
$m_D=0.5$ GeV (upper panel) and $m_D=2.0$ GeV (lower panel).
In both panels we compare the Chapman-Enskog result (dashed red line)
with the Relaxation time result (dot-dashed blue line).}
\end{figure}
In Fig. \ref{Fig:5abMD} we plot $P_L/P_T$ as a function of time 
for two values of $m_D$: in the upper panel we consider $m_D = 0.5$ GeV
and in the lower panel $m_D = 2$ GeV. In each panel, the red dashed line corresponds
to the CE formula while the dot dashed blue line to the RTA formula.
In the case of $m_D=0.5$ GeV 
the oscillations of $P_L/P_T$ with CE are quite damped with respect to those
of the RTA. This behaviour is easily understood by the results of \cite{Plumari:2012ep}:
as a matter of fact a given $\eta/s$ corresponds to a lower cross section for RTA,
implying larger conductive currents which sustain plasma oscillations.
In the case of $m_D = 2.0$ GeV the behaviour of $P_L/P_T$ in RTA is similar
to that in CE; in fact in this case the cross section is mostly isotropic, and it is known
that RTA cross section is quite close to the CE one.

\section{Summary and Conclusions}
In this article we have reported about our results on the simulations of the Schwinger
effect for color electric flux tubes, focusing on thermalization and isotropization
of the fluid produced by the decay of the electric field in the tube.
These simulations are important in the context of ultrarelativistic nuclear collisions,
where flux tubes of strong color fields are expected to be produced
in the early stage.
According to the general understanding of high energy collision processes,
this work is relevant both for heavy ion collisions, where a large number of tubes is expected,
and  for proton-nucleus and proton-proton collisions.

We have studied both the cases of flux tube decay in a static box and 
in a box with a longitudinal expansion. 
The flux tube decay is described by the Schwinger effect;
the dynamical evolution of the field is assumed to be abelian so it satisfies
Maxwell equations, and the 
dynamics of the fluid produced by the decay is studied by the relativistic
transport theory coupled to the Maxwell equations.
We have formulated transport theory in terms of a fixed value of $\eta/s$
rather than scattering amplitude; in this way we are able to describe the 
evolution of the fluid in terms of macroscopic quantities ($\eta/s$)
rather than insisting on a set of specific microscopic processes.
The approach developed here, based on a stochastic solution of the relativistic
Boltzman equation, has in perspective the advantage to be easily extended to 
$3+1$D realistic simulations, allowing to study the possible impact of the early dynamics on observables
like radial and elliptic flow~\cite{prepa}.
Our study well matches \cite{Ryblewski:2013eja} where the same problem has been studied.
The main difference between the present study and \cite{Ryblewski:2013eja} is that in the latter
the relaxation time approximation is used to simplify the collision integral
in the Boltzmann equation, while in our study we do not use such an approximation
hence solving the relativistic kinetic equation with the full collision integral.

We have considered the case of a system with boost invariant longitudinal expansion,
which is relevant for the case of simulation of early stages in 
heavy ion collisions.
We have studied many aspects characterizing time evolution of the flux tube.
In particular we have computed decay time of the color electric field, 
which we find to be a small fraction of fm/c; for large values of $\eta/s$
we have found plasma oscillations develop within $1$ fm/c
and persist for several fm/c in agreement with \cite{Ryblewski:2013eja}; 
on the other hand for $\eta/s\leq0.3$  plasma oscillations are very mild
or even absent.
 
We have also computed the time evolution of the longitudinal to the transverse
pressure ratio $P_L/P_T$. Given the particular initial field configuration and the absence
of particle quanta at $\tau=0$ we have $P_L/P_T=-1$ at initial time; however longitudinal pressure
becomes positive in $\tau\approx0.2$ fm/c because particles are produced by the Schwinger mechanism,
giving a positive contribution to the pressure.
For $\eta/s\leq0.3$ we have found isotropization
times $\tau_{iso}\leq$ 0.8 fm/c; for larger values of $\eta/s$
 plasma oscillations cause the system to be  less isotropic.
Our results imply that for a fluid with $4\pi\eta/s\leq3$ 
isotropization time is less than $1$ fm/c. This would justify
the use of viscous hydrodynamics with initial times $\tau_0\approx0.6\div0.8$ fm/c
for $4\pi\eta/s=1\div 3$. It remains to be understood if the oscillations of $P_L/P_T$
for $t\leq1$ fm/c affect some observable in ultrarelativistic heavy ion collisions.
Our approach based on the stochastic solution of the Boltzmann equation is already set up
to perform simulations starting from more realistic initial conditions 
and evolutions in 3+1 dimensions, which will be useful to understand quantitatively 
the effect of such early times behaviour of physical quantities on observables.

We have also studied proper energy density and temperature evolution
as well as particle spectra in the transverse
plane for several values of $\eta/s$.
We have found that in case of small $\eta/s$ the dynamics is very efficient
in both removing the initial anisotropy and in getting a quick thermalization;
in fact for $4\pi\eta/s=1$ we find thermalization time $\tau_{therm}\leq1$ fm/c
by comparing the particle spectrum at $t=1$ fm/c with the thermal spectrum
Eq.~\eqref{eq:thermalspe} where we put the temperature equal to 
the one we obtain in the simulation, see lower panel of Fig.~\ref{Fig:2aNEW}.
On the other hand, increasing $\eta/s$ results in a mismatch between the thermal
and the actual spectra at $t=1$ fm/c, meaning thermalization time is larger in these cases.

To check the effect of microscopic details on the quantities we have computed,
we have turned the microscopic cross section from isotropic to anisotropic
by changing the value of the Debye screening mass in the two body cross section 
while keeping $\eta/s$ fixed. 
We have found that changing $m_D$ does not lead to significant
changes in the evolution of physical quantities, 
besides small deviations in the early times evolution and
some mild damping of field oscillations in the case of the largest $\eta/s$.

We have also considered the effect of turning from the  CE
to Relaxation Time Approximation (RTA) formula connecting
the shear viscosity to the total cross section,
still keeping the full collision integral in the Boltzmann equation.
This is not exactly what has been done in \cite{Ryblewski:2013eja}
where RTA is used as an ansatz for the collision integral.
We have found that in the case of a forward peaked cross section
changing from CE to RTA affects the behaviour of $P_L/P_T$,
increasing oscillations at early times. Nevertheless late times dynamics
is almost unaffected by this change.

There are several points we have not considered in this study.
We have neglected rapidity fluctuations in the electric field
which would break boost invariance. 
Moreover in our description we have 
not included a longitudinal
magnetic field at initial time which might modify sligthly the Schwinger amplitude \cite{Tanji:2008ku}.
Both fluctuations in rapidity and the initial chromomagnetic field 
are quite interesting ingredients to add to the present study and we plan to 
add them in future works. Moreover it would be of a certain interest 
to consider geometries with several flux tubes 
pinned on the transverse plane, hence pushing our calculations
towards more realistic initial state conditions for
heavy ion collisions as well as 
quantitative predictions of observables of phenomenological interest like multiplicities,
particle spectra and collective flows. We will consider these developments
in forthcoming publications.

\emph{Acknowledgements.} 
V. G., S. P., M. R. and F. S. acknowledge the ERC-STG funding under the QGPDyn
grant. M. R. acknowledges discussions with T. Epelbaum, W. Florkowski and R. Ryblewski.

  \end{document}